%% file: g-2CNH.tex
\documentclass[12pt]{article}
\usepackage{epsfig,cite,feynarts,amsfonts,amsmath,amssymb,psfrag}

\input paperdef

\begin{document}
\thispagestyle{empty}

\def\thefootnote{\fnsymbol{footnote}}

\begin{flushright}
CERN--PH--TH/2004-094\\
DCPT/04/44\\
IPPP/04/22\\
hep-ph/0405255
\end{flushright}

\vspace{1cm}

\begin{center}

{\Large\sc {\bf Electroweak and supersymmetric}}

\vspace{0.4cm}

{\Large\sc {\bf two-loop corrections to $(g-2)_\mu$}}

\vspace{1cm}

{\sc 
S.~Heinemeyer$^{1}$%
\footnote{email: Sven.Heinemeyer@cern.ch}%
, D.~St\"ockinger$^{2}$%
\footnote{email: Dominik.Stockinger@durham.ac.uk}%
~and G.~Weiglein$^{2}$%
\footnote{email: Georg.Weiglein@durham.ac.uk}
}

\vspace*{1cm}

{\sl
$^1$ CERN, TH Division, Dept. of Physics, 1211 Geneva 23, Switzerland

\vspace*{0.4cm}

$^2$Institute for Particle Physics Phenomenology, University of Durham,\\
Durham DH1~3LE, UK
}

\end{center}

\vspace*{0.2cm}

\begin{abstract}

We present the up to now most precise evaluation of electroweak and
super\-symmetric contributions to the anomalous magnetic moment of the
muon $\amu$, describing in detail also the calculational
techniques. We calculate the bosonic two-loop contributions in the
Standard Model without the approximation of a heavy Higgs-boson mass,
finding corrections up to $0.2\times10^{-10}$ for a light Higgs
boson. In the Minimal Supersymmetric Standard Model the corresponding
two-loop contributions from the two-Higgs-doublet model part differ from 
the Standard Model result
by up to $0.3\times10^{-10}$.
Finally, we evaluate the diagrams where a loop of charginos or
neutralinos, the superpartners of gauge and Higgs bosons, is inserted
into a two-Higgs-doublet one-loop diagram. These corrections can amount up
to $10\times10^{-10}$, which is almost $2\si$ of the current
experimental uncertainty.
\end{abstract}

\def\thefootnote{\arabic{footnote}}
\setcounter{page}{0}
\setcounter{footnote}{0}

\newpage


\section{Introduction}

The final result of the Brookhaven ``Muon $g-2$ Experiment'' (E821) for
the anomalous magnetic moment of the muon, $\amu \equiv (g-2)_\mu/2$, 
reads \cite{g-2exp}
\BE
\amuexp = (11\, 659\, 208 \pm 6) \times 10^{-10}~.
\label{eq:amuexp}
\EE
The Standard Model (SM) prediction depends on the evaluation of the
hadronic vacuum polarization and light-by-light contributions. The
former have been evaluated by \citeres{DEHZ,g-2HMNT,Jegerlehner,Yndurain}, 
the latter by \citere{LBL}, but there is a recent
reevaluation~\cite{LBLnew}, describing a possible shift of the central
value by $5.6 \times 10^{-10}$.
Depending on which hadronic evaluation is chosen, the
difference between experiment and the SM prediction lies between the two
values%
\footnote{
We always include the updated QED result from \citere{Kinoshita}.
}%
\BEA
\label{deviation1}
\amuexp-\amutheo ~(\mbox{\cite{g-2HMNT}+\cite{LBL}}) & = &
(31.7\pm9.5)\times10^{-10} ~:~3.3\,\si~,\\
\amuexp-\amutheo ~(\mbox{\cite{DEHZ}+\cite{LBLnew}}) & = &
(20.2\pm9.0)\times10^{-10} ~:~2.1\,\si~.
\label{deviation2}
\EEA
These evaluations are all $e^+e^-$ data driven. 
Recent analyses concerning $\tau$ data indicate that uncertainties due to
isospin breaking effects may have been underestimated
earlier~\cite{Jegerlehner}.
For the purpose of numerical comparisons we will use the intermediate
value
\BEA
\label{deviationfinal}
\amuexp-\amutheo ~(\mbox{\cite{g-2HMNT}+\cite{LBLnew}}) & = &
(24.5\pm9.0)\times10^{-10} ~:~2.7\,\si~.
\EEA

It is an interesting question
whether the observed $2-3\si$ deviation is due to supersymmetric
effects. The supersymmetric one-loop contribution \cite{g-2MSSMf1l}
is approximately given by 
\BE
\amu^{\SU,{\rm 1L}} = 13 \times 10^{-10} 
             \KL \frac{100 \gev}{\msusy} \KR^2 \tb\;
 {\rm  sign}(\mu),
\label{susy1loop}
\EE
if all supersymmetric particles (the relevant ones are the smuon,
sneutralino, chargino and neutralino) have a common mass
$\msusy$. Obviously, supersymmetric effects can easily account for a
$(20\ldots30)\times10^{-10}$ deviation, if $\mu$ is positive and
$\msusy$ lies roughly between 100 GeV (for small $\tb$) and
600 GeV (for large $\tb$). This mass range is both allowed by
present search limits and very interesting in view of physics at Run~II
of the Tevatron, the LHC and a future Linear Collider (LC).

Eq.\ (\ref{susy1loop}) also shows that for certain parameter choices
the supersymmetric contributions could have values of either
$\amu^{\SU}\gsim60\times10^{-10}$ or $\amu^{\SU}\lsim-10\times 10^{-10}$,
both outside the $3\si$ band of the allowed range according to
(\ref{deviation1}), (\ref{deviation2}). This means
that the $(g-2)_\mu$ measurement places strong bounds on the supersymmetric
parameter space. This is important for constraining different variants
of supersymmetric models (and of course also other models of new
physics) and complements the direct searches. 
Even after
the discovery of supersymmetric particles, indirect bounds derived from 
$(g-2)_\mu$ or $b$-decays and a Higgs mass measurement will provide
important complementary information to that obtained from direct measurements.

In order to fully exploit the precision of the $(g-2)_\mu$ experiment 
within Supersymmetry (SUSY), see 
\citeres{g-2appl1,g-2appl2} for possible applications, the theoretical
uncertainty of the SUSY loop contributions from unknown higher-order
corrections should be significantly lower than the experimental error
given in \refeq{eq:amuexp} and the hadronic
uncertainties in the SM
prediction. Thus, the reduction of the uncertainty of the SUSY loop
contributions down to the level of about $\pm1\times10^{-10}$ is
desirable. 

For the electroweak part of the SM prediction this accuracy has been
reached with the computation of the complete two-loop 
result~\cite{g-2SM2lA,g-2SM2lB}. However, this result relies on a
single evaluation of the bosonic two-loop contributions, performed in
the limit of a heavy SM Higgs-boson mass, $\MHSM \gg \MW$. Thus, an
independent check seems 
desirable. In this paper we perform this evaluation for arbitrary
Higgs-boson masses. 

For the SUSY contributions, a similar level of accuracy has not been
reached yet, since the corresponding
two-loop corrections are 
largely unknown. Only two parts of the two-loop contribution have been
evaluated up to now. The
first part are the leading $\log \KL m_\mu/\msusy\KR$-terms of
supersymmetric one-loop diagrams with a photon in the second loop. They
amount to about $-8\%$ of the supersymmetric one-loop contribution
(for a SUSY mass scale of $\msusy = 500
\gev$)~\cite{g-2MSSMlog2l}. 

The second known part are the diagrams
with a closed loop of SM fermions or scalar fermions calculated in
\citere{g-2FSf}, extending previous results of
\citeres{g-2BarrZee1,g-2BarrZee2}. 
It has been shown in \citere{g-2FSf} that the numerical effect of these
contributions could in principle be as large as $20\times10^{-10}$ for
suitable parameter choices. But experimental 
constraints on the lightest Higgs-boson
mass~\cite{mhiggslong,mhiggsAEC,LEPHiggsSM,LEPHiggsMSSM}, electroweak
precision observables~\cite{delrhosusy1loop,delrhosusy2loop,delrho2lC} and
$b$-decays~\cite{bsmumu,bsg} restrict the 
allowed parameter space. Thus the values of these diagrams amount up
to about $5\times10^{-10}$, except in rather restricted parameter
regions with non-universal sfermion mass parameters involving very
disparate mass scales.

In general, every diagram in the Minimal Supersymmetric
Standard Model (MSSM) must contain one continuous line carrying the
$\mu$-lepton number. Thus, the MSSM diagrams can be divided into two
classes: (1) diagrams 
containing a one-loop diagram involving supersymmetric particles
(with a $\tilde\mu$ or
$\tilde\nu_\mu$ line) to which a second loop is attached, and
(2) diagrams where a second loop is attached to a two-Higgs-doublet
model one-loop diagram (with a $\mu$ or $\nu_\mu$ line). The
QED-logarithms from \citere{g-2MSSMlog2l} belong to the first class, and
the fermion/sfermion two-loop diagrams from \citere{g-2FSf} to the second.

The diagrams of the second class are particularly interesting
since they can depend on other parameters than the supersymmetric
one-loop diagrams and can therefore change the qualitative behaviour
of the supersymmetric contribution to $\amu$. In particular, they could
even be large if the one-loop contribution is suppressed, e.g.\ due to
heavy smuons and sneutrinos.

In this paper we complete the calculation of the diagrams of this
class. 
As the first step we calculate all pure two-Higgs-doublet
model diagrams. This calculation is analogous to the one of the SM
bosonic two-loop corrections (we carry out the calculation of the SM
contributions as a special case), but it involves the additional Higgs
bosons (but no SUSY particles). 
The second step comprises the diagrams with a closed
chargino/neutralino loop. Their structure is similar to the one of
diagrams with a closed SM fermion or sfermion loop, but they depend on a
completely different set of parameters and show a complementary
behavior.

The objective of this paper is to describe in detail the calculational
steps, which have already been used in \citere{g-2FSf}, and to
analyze the numerical impact of the newly derived SUSY contributions. 
The SM result is compared to the existing calculation.
As for the results of \citere{g-2FSf}, the new results will be
implemented into the Fortran code \fh~\cite{feynhiggs}.

The outline is as follows. In Sect.\ 2 we describe the calculation, in 
particular the
used regularization, large mass expansion, reduction of 
two-loop integrals, and renormalization. For the two-loop QED
corrections involving SUSY particles analytical results are presented. 
Section 3 is devoted to the SM contributions. The 
numerical results for the two-Higgs-doublet contribution in the MSSM and
the chargino/neutralino contribution
are discussed in Sects.~4 and 5, respectively. Our conclusions are
given in Sect.~6.


\section{Calculation}
\label{sec:calc}

\subsection{Extraction of $\amu$}

The anomalous magnetic moment $\amu$ of the muon is related to the 
photon--muon vertex function $\Gamma_{\mu\bar\mu A^\rho}$ as follows:
\BEA
\label{covdecomp}
\bar u(p')\Ga_{\mu\bar\mu A^\rho}(p,-p',q) u(p) & = &
\bar u(p')\left[\ga_\rho F_V(q^2) + (p+p')_\rho F_M(q^2) +
  \ldots\right] u(p),\\
\amu & = & -2m_\mu F_M(0).
\EEA
It can be extracted from the regularized vertex function using the
projector~\cite{g-2SM2lA,g-2SM2lB}
\BEA
\label{projector}
\amu & = & \frac{1}{2(D-1)(D-2)m_\mu^2} {\rm Tr}\Bigg\{
            \frac{D-2}{2}\left[m_\mu^2\ga_\rho - D p_\rho\pslash -
            (D-1)m_\mu p_\rho\right]V^\rho\nonumber\\
&&\qquad\quad +\frac{m_\mu}{4}
  \left(\pslash+m_\mu\right)
\left(\ga_\nu\ga_\rho-\ga_\rho\ga_\nu\right)
  \left(\pslash+m_\mu\right)T^{\rho\nu}\Bigg\},\\
V_\rho & = & \Ga_{\mu\bar\mu A^\rho}(p,-p,0),\\
T_{\rho\nu} & = & \frac{\partial}{\partial q^\rho}
\Ga_{\mu\bar\mu A^\nu}(p-(q/2),-p-(q/2),q)\bigg|_{q=0}.
\EEA
Here the muon momentum is on-shell, $p^2=m_\mu^2$, and $D$ is the
dimension of space-time.

This projector requires that in the covariant decomposition
(\ref{covdecomp}) only $D$-di\-men\-sional quantities appear. We
therefore use dimensional regularization with anti-commuting
$\ga_5$, where there is no distinction between the first four and
the remaining $D-4$ dimensions. However, in order to demonstrate the
validity of this
regularization we need to discuss three issues related to
supersymmetry breaking, mathematical consistency and the
treatment of the $\epsilon$-tensor in this scheme.
\begin{itemize}
\item Dimensional regularization breaks supersymmetry
  \cite{susybreaking}, and in general 
  supersymmetry has to be restored by adding certain counterterms not
  corresponding to multiplicative renormalization of the fields
  and parameters \cite{susyrestore}. In our case, however, all
  appearing counterterms  are two-Higgs-doublet model counterterms,
  because we calculate 2-loop corrections to 1-loop two-Higgs-doublet
  model diagrams, see also \refse{subsec:ct}. 
  These two-Higgs-doublet model counterterms are either
  fixed by renormalization conditions, like the muon- or $Z$-mass
  counterterm, or by gauge invariance, like the $\mu\mu\ga$- or
  $W^+G^-\ga$-counterterms. Since gauge invariance is not broken by
  dimensional regularization with anti-commuting $\ga_5$,%
\footnote{%
This has been checked up to the order we need here in
  \citere{mudec}.
}%
~multiplicative renormalization is sufficient for
  all counterterms in our calculation. 
\item Using an anti-commuting $\ga_5$ in $D\ne4$ implies ${\rm Tr}(\ga_5
  \ga^\mu\ga^\nu\ga^\rho\ga^\si)=0$ and is therefore
  incompatible with the trace formula ${\rm Tr}(\ga_5
  \ga^\mu\ga^\nu\ga^\rho\ga^\si)\propto
  \epsilon^{\mu\nu\rho\si}$, which has to be reproduced for
  $D\to4$. As it is customary, we simply replace each trace of the
  form ${\rm Tr}(\ga_5 \ga^\mu\ga^\nu\ga^\rho\ga^\si)$ by its
  four-dimensional value. We can check the correctness of this
  procedure by comparing the terms involving $\epsilon$-tensors with
  the ones obtained using the HVBM-scheme \cite{HVBM}, which is fully
  consistent. In our calculation, $\epsilon$-tensors appear only in
  two-loop diagrams with an insertion of a fermion triangle with three
  external vector bosons. The difference of the fermion triangle
  subdiagrams in the two schemes (summed over one
generation or all charginos and neutralinos) is of the order
  $(D-4)\frac{k_\si}{k^2}\,\epsilon^{\mu\nu\rho\si}$, where $k_\si$ is
the non-vanishing external momentum of the fermion triangles. 
The covariant $\epsilon^{\mu\nu\rho\si}\,k_\si$ of power-counting
degree $+1$ has the same prefactor as the chiral gauge anomaly and
thus vanishes.
  If the fermion triangles are inserted into 1-loop diagrams contributing
  to   $\Ga_{\mu\bar\mu A^\rho}$, the difference between the naive
  and the HVBM-scheme vanishes for $D\to4$ for power-counting
reasons. Hence, the terms involving 
  $\epsilon$-tensors are identical in the naive and the HVBM-scheme,
  which means that the naive scheme produces these terms correctly (see
  also the discussion in \citere{mudec}).
\item The $\epsilon$-tensor is a purely four-dimensional object, and
  contractions like 
  $\epsilon^{\mu\al\be\ga}\epsilon^{\nu}{}_{\al\be\ga}$ or
  $\epsilon^{\mu\nu\al\be}\epsilon^{\rho\si}{}_{\al\be}$
  have to be evaluated in four dimensions.%
\footnote{%
We checked explicitly that a $D$-dimensional treatment
indeed leads to incorrect results.
}%
  ~Effectively, this leads to the appearance of the additional covariant 
$\ga_\rho^{\rm 4-dim} F_V^{\rm 4-dim}$ in \refeq{covdecomp}. 
  This contradicts the requirement of the projection
  (\ref{projector}) that only $D$-dimensional covariants may appear in
  $\Ga_{\mu\bar\mu A^\rho}$. However, as mentioned above, the
  fermion triangle subdiagrams add up to an expression of reduced
  power-counting degree, and accordingly the $\epsilon$-tensor
  contributions to $\Ga_{\mu\bar\mu A^\rho}$ and thus $F_V^{\rm 4-dim}$ 
are finite. 
As a result, the projection operator (\ref{projector}) produces the
  correct result, containing only the form factor $F_M$, even if the
$\epsilon$-tensors are evaluated in four dimensions.
\end{itemize}


\subsection{Diagram evaluation}
\label{subsec:diageval}

The actual calculation is done using computer algebra. The 
Feynman diagrams are generated using {\em  FeynArts}
\cite{feynarts,fa-mssm}. After applying the projector
(\ref{projector}), the Dirac algebra, traces and contractions are
performed by {\em TwoCalc} \cite{2lred}. 

The main part of the two-loop calculation consists of the evaluation of
the two-loop integrals and the simplification of their
coefficients, both of which is complicated by the large number of
different mass scales and the involved structure of the MSSM Feynman
rules. 

As a first step we perform a large mass expansion \cite{smirnov} in
the ratio $m_\mu/M_{\rm heavy}$, where $M_{\rm heavy}$ stands for all
heavy masses, $M_{Z,W}$ and Higgs and chargino/neutralino masses. 
Depending on the prefactors we expand all integrals to sufficiently
high order such that the final result is correct up to 
\order{m_\mu^2/M_{\rm heavy}^2}. We discard all terms of
  \order{m_\mu^4/M_{\rm heavy}^4}. The expansion is based on the
formula \cite{smirnov}
\BEA
F_\Ga \sim \sum_{\ga} F_{\Ga/\ga}\ \circ\ {\cal
  T}_\ga\;F_\ga.
\label{LMExp}
\EEA
Here $F_\Ga$ denotes the integral corresponding to a (two-loop) Feynman
diagram $\Ga$, $\Ga/\ga$ denotes the diagram where the
subdiagram $\ga$ is shrunk to a point and ${\cal T}_\ga$ denotes
Taylor expansion with respect to all small masses and all external
momenta of $\ga$. The sum runs over all (in general unconnected)
subdiagrams $\ga$ that 
\begin{itemize}
\item contain all lines with heavy masses,
\item are one-particle irreducible with respect to the light lines.
\end{itemize}
By means of (\ref{LMExp}) the large masses appear only in $\ga$,
whereas the small masses and momenta appear only in $\Ga/\ga$,
resulting in a separation of scales.
Since the projection operator (\ref{projector}) sets the external
photon momentum to zero, our integrals are initially two-loop
two-point integrals. There are therefore three possibilities for the
r.h.s.\ of the large mass expansion~(\ref{LMExp}) and the subsequent
reduction to master integrals: 
\begin{itemize}
\item \ul{(Light 0-loop) $\circ$ (heavy 2-loop):} 
  In this case, the r.h.s.\ of
  (\ref{LMExp}) results in a two-loop vacuum diagram times a rational
  function of $m_\mu$. Using partial integration
  identities~\cite{pIId}, we reduce all two-loop vacuum diagrams to
  the master integral
\BEA
T_{134}(m_1,m_3,m_4) & = &
\langle\langle\frac{1}{D_1D_3D_4}\rangle\rangle
\EEA
in the notation of \cite{2lred}:
\BEA
D_i& = & k_i^2-m_i^2,\\
k_1 & = & q_1,\ k_2=q_1+p,\ k_3=q_2-q_1,\\
k_4 & = & q_2,\ k_5=q_2+p,\\
\langle\langle\ldots\rangle\rangle & = & 
\int\frac{d^Dq_1\;d^Dq_2}{[i\pi^2(2\pi\mu)^{D-4}]^2}(\ldots).
\EEA
The result for the general case of three different masses can be found
e.g.\ in \citere{t134}.
\item \ul{(Light 1-loop) $\circ$ (heavy 1-loop):} 
  In this case the integrals
  can be reduced to the standard one-loop functions $A_0(m)$
  and $B_0(m_\mu^2,0,m_\mu)$~\cite{a0b0c0d0}.
\item \ul{(Light 2-loop) $\circ$ (heavy 0-loop):} 
  This case appears e.g.\ in the calculation 
  of diagrams~7,8 in \reffi{fig:logMHdiag} below. In general it can only
  appear in diagrams that do not involve any SUSY particles, 
  since SUSY particles (in scenarios with R-parity
  conservation) necessarily form at least one closed
  heavy loop. The r.h.s.\ of (\ref{LMExp}) then contains 
  two-loop two-point functions, however with only one mass scale,
  $m_\mu$. One typical example of such integrals is
\BEA
Y^{1155}_{2334}(m_\mu^2; m_\mu, m_\mu, m_\mu) &  = &
\langle\langle\frac{(k_1^2)^2(k_5^2)^2}{D_2D_3^2D_4}\rangle\rangle.
\label{Yexample}
\EEA
with $p^2=m_\mu^2$ and $m_{2,3,4}=m_\mu$. Using Passarino-Veltman
decomposition for the one-loop subdiagrams~\cite{2lred} and partial 
integration identities all such integrals can be reduced to one-loop 
integrals and the on-shell (i.e.\ $p^2=m_\mu^2$)
master two-loop sunset integrals $T_{234}(m_\mu^2;m_\mu,m_\mu,m_\mu)$ and
$T_{234}(m_\mu^2;m_\mu,0,0)$. For example, for
(\ref{Yexample}) we obtain 
\BEA
\lefteqn{Y^{1155}_{2334}(m_\mu^2; m_\mu, m_\mu, m_\mu)  = }
\nonumber\\
&&\frac{4(-4 + 19D - 31D^2 + 14D^3)}{8 - 18D + 9D^2}
\;  m_\mu^4 A_0(m_\mu)^2 \\
&&{}
 + \frac{32 (-2 + D)D^2}{3(8 - 18D + 9D^2)}
\;  m_\mu^6T_{234}(m_\mu^2;m_\mu,m_\mu,m_\mu)~. \nonumber
\EEA
For the analytical results of on-shell two-loop two-point functions,
see \citere{onshell2} and references therein. Note that the reduction
can generate spurious additional divergences, in the form of
$1/(D-4)$-poles in prefactors of master integrals. A simple
example is given by the reduction of $Y^{1}_{2345}$ with 
$p^2=m_\mu^2$ and $m_{2,3,4}=m_\mu$, $m_5=0$:
\BEA
\lefteqn{Y^{1}_{2345}(m_\mu^2; m_\mu, m_\mu, m_\mu, 0) = } \non \\
&&-\frac{1}{2(D-4)}\bigg\{
 4 A_0(m_\mu)\; B_0(m_\mu^2, 0, m_\mu)
\nonumber\\
&&{}\quad + (D-2 )\bigg[\frac{A_0(m_\mu)^2}{m_\mu^2(D-3)} 
 - 2 T_{234}(m_\mu^2;m_\mu, m_\mu, m_\mu)\bigg]\bigg\}.
\EEA
Although the $1/(D-4)^3$-poles cancel, the
analytical results of the master integrals are needed up to 
\order{D-4} in the case of $T_{234}$ and up to \order{(D-4)^2} in
the case of $A_0$ and $B_0$ functions.
\end{itemize}


\subsection{Counterterm contributions}
\label{subsec:ct}

Since $\amu$ is exactly zero at tree-level, no two-loop
counterterm corrections arise. However, one-loop diagrams with
counterterm insertion are necessary to derive a UV-finite result. 
Thus, together with the two-loop diagrams we have also evaluated all the
diagrams with subloop renormalization in the SM and the MSSM.

In order to derive the amplitudes of the diagrams with counterterm
insertion it was necessary to extend the existing \fa\ MSSM model
file~\cite{fa-mssm} by including non-SM counterterm
vertices. The introduced counterterms contain field
renormalization constants for the gauge and Higgs bosons and for the
muon and the neutrino. We have checked that in the sum of all
amplitudes all field renormalization constants, except of the external
muon and photon, drop out as required. In addition to the field
renormalization constants also mass counterterms for the SM gauge
bosons and the muon have been introduced. Finally, 
counterterms for the Higgs sector arising from the Higgs potential (see
e.g.\ \citere{mhiggslong}) as well as for the mixing of Higgs-
and gauge bosons have been included. The Higgs sector counterterms
consist, besides 
field renormalization constants, of Higgs tadpoles that cancel large
corrections coming from the self-energy insertions at the two-loop
level. 

We have chosen on-shell renormalization conditions for the muon and the
photon, for the $W$- and $Z$-boson masses, and for the electric charge.
The tadpole counterterms cancel their corresponding loop contributions.

Since all amplitudes are reduced to self-energy like
corrections, the result of the counterterm diagrams consists only of
$A_0$ and $B_0$ functions. 
As for the genuine two-loop diagrams, also for the counterterm diagrams 
a large mass expansion is necessary 
in order to obtain a consistent expansion in
powers of $m_\mu$. 

\vspace{-.2em}
As already outlined in \citere{g-2FSf} we parametrize our one-loop
result with $\gf$ in order to absorb process-independent
higher-order corrections. This requires the evaluation of the one-loop
corrections to muon decay, $\De r$, from
the corresponding sectors. For the class of corrections evaluated 
in \citere{g-2FSf},
all contributions to $\De r$ from scalar fermion
diagrams (since the SM fermions contribute only to the SM result) had to
be taken into account. For
the results computed in the present paper the $\De r$ corrections 
from charginos and neutralinos, as well as from the whole
two-Higgs-doublet sector of the MSSM (except fermions) had to be
evaluated. While the former ones consist only of self-energy
corrections, the latter ones also contain vertex and box contributions
involving SM particles. The $\De r$ contribution enters via
$(- a_{\mu}^{\rm 1 L} \, \Delta r)$, where $a_{\mu}^{\rm 1 L}$ is the
two-Higgs-doublet model one-loop result. In the case of the SM gauge and
Higgs-boson contributions the $\Delta r$ term amounts to a shift in the
two-loop corrections of about 10\%.


\subsection{QED contributions}

In general, the contributions to $\amu$ can be split into QED and
electroweak contributions. A diagram is counted as a QED diagram if it
involves only the photon but no other gauge or Higgs boson, and as an
electroweak diagram if it contains a $W$, $Z$ or Higgs boson (or some
of the corresponding ghost fields). 

At the two-loop level, there are QED diagrams involving a photon
vacuum polarization subdiagram (see \reffi{fig:QEDgeneral}). For
leptons or quarks, these contributions are of course known; in the
latter case they have to be obtained via the hadronic vacuum
polarization~\cite{DEHZ,g-2HMNT,Jegerlehner,Yndurain}.

\input{QEDdiagrams.tex}

If supersymmetric particles contribute, the two-loop QED contributions are
modified by the diagrams of \reffi{fig:QEDgeneral} with a
chargino, slepton, or squark loop. In our calculations in the
following sections, as well as in \citere{g-2FSf}, we do not include
these QED contributions. They can be easily inferred from known
results~\cite{refsQED} and are tiny. For completeness, we list the
results here. 
The result of these diagrams depends only on the charge $Q$, the mass
$M$, and the spin of the particle in the vacuum polarization. For a
scalar particle we have
\BEA
\amu^{\rm QED,2L,scalar} & = &
\frac{Q^2}{360}\left(\frac{\al}{\pi}\right)^2\frac{m_\mu^2}{M^2}~, 
\label{eq:qed1}
\EEA
and for a fermionic particle we have
\BEA
\amu^{\rm QED,2L,fermionic} & = & 
\frac{Q^2}{45}\left(\frac{\al}{\pi}\right)^2\frac{m_\mu^2}{M^2}~.
\label{eq:qed2}
\EEA
For masses $M\gsim100\gev$, these contributions are below $10^{-13}$
and hence negligible.

The results in (\ref{eq:qed1}), (\ref{eq:qed2}) 
contain the suppression factor $m_\mu^2/M^2$, which is
typical for all diagrams except for the pure QED-ones involving only
photons and muons. However, this only holds after
renormalization,
i.e.\ for the sum of the two-loop diagrams and the counterterm
insertions. The two-loop diagram in \reffi{fig:QEDgeneral} itself has
a contribution of the order $m_\mu^0$, which is then cancelled by the
counterterm diagram. The reason is that before renormalization, the
vacuum polarization subdiagram including the two photon propagators
behaves like $\frac{1}{k^2}\Pi^\gamma(k^2)\frac{1}{k^2}\propto
\frac{1}{k^2}$ for $k^2 \to 0$, like the tree-level photon propagator, 
whereas after renormalization 
(here charge and field renormalization effectively amount to adding the
counterterm diagram in \reffi{fig:QEDgeneral})
we have 
$\frac{1}{k^2}\hat\Pi^\gamma(k^2)\frac{1}{k^2}\propto1$ for $k^2\to0$.


\section{Standard Model contributions}

Before discussing the MSSM results, we present our results for the
bosonic electroweak two-loop contributions in the SM.
Up to now, there exists only one evaluation of these diagrams
\cite{g-2SM2lA}, which employs the approximation $\MHSM \gg \MW$. Our
recalculation of these contributions serves both
as a cross check of \citere{g-2SM2lA} and of our algebraic codes. It
furthermore allows to compare the approximation of \citere{g-2SM2lA} with
the result for arbitrary $\MHSM$.

\begin{figure}[htb!]
\begin{center}
\epsfig{figure=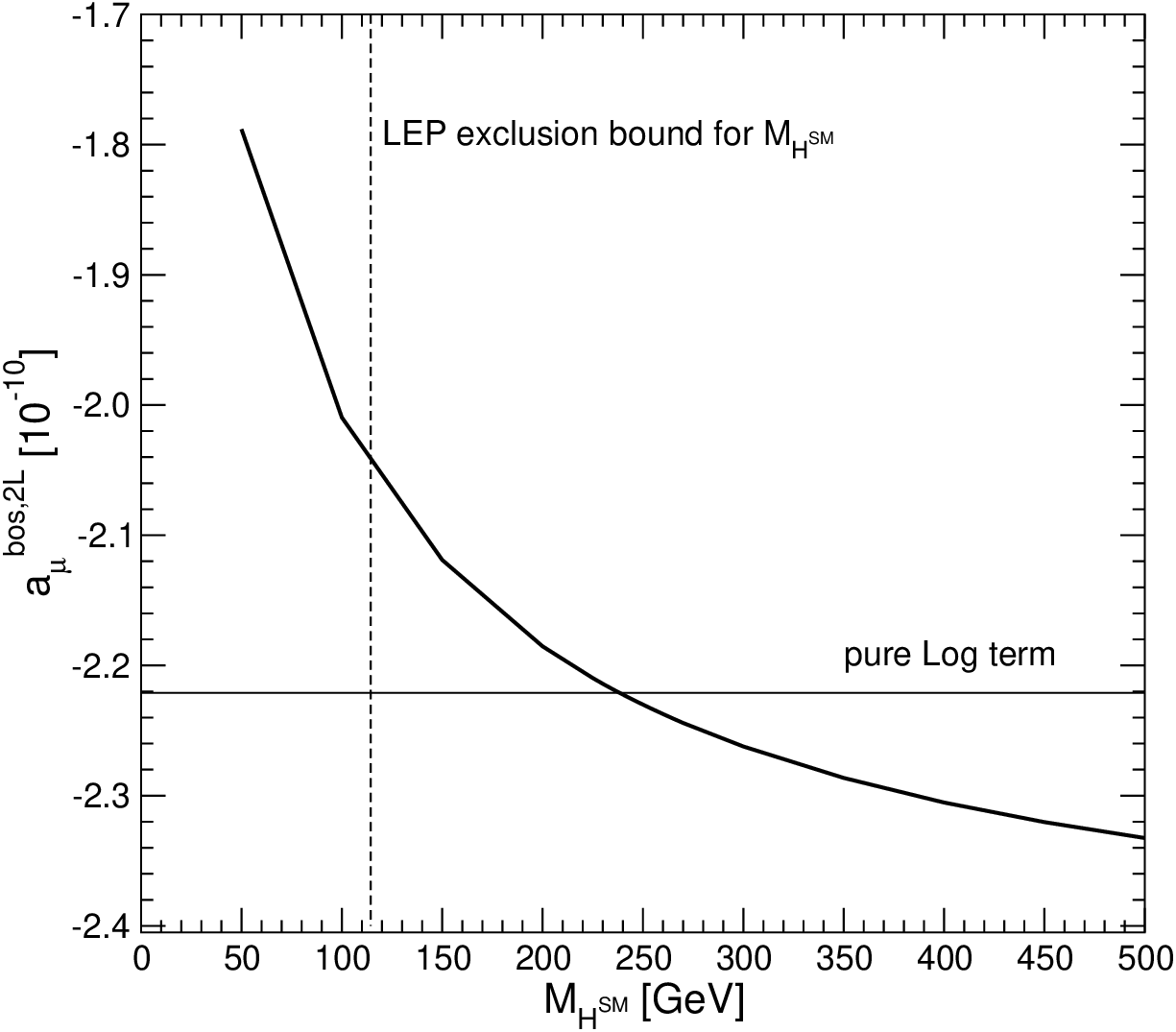,width=12cm,height=8cm}
\caption{%
$\amu^{\rm bos,2L}$ is shown as a function of $\MHSM$, see
  \refeq{resSMbosonic}. The size of the (constant) logarithmic
  contributions is also indicated. 
}
\label{fig:SMbosonic}
\end{center}
\end{figure}

Some typical diagrams of this class are shown in
\reffi{fig:logMHdiag} below, if one identifies $\phi$ with the 
SM Higgs boson and $\psi$ with the charged Goldstone boson of the SM (in the
Feynman gauge). 
In general this class consists of all SM
two-loop diagrams without a closed fermion loop and without pure QED
diagrams. We obtain the following result:
\BEA
\amu^{\rm bos,2L} & = & \frac{5}{3}\frac{\gf m_\mu^2}{8 \pi^2
  \sqrt{2}} \;\frac{\al}{\pi}
\KL c^{\rm bos,2L}_L \log\frac{m_\mu^2}{M_W^2} 
+ c^{\rm bos,2L}_0 \KR,
\label{resSMbosonic}
\EEA
where the coefficient of the large logarithm,
$\log\frac{m_\mu^2}{M_W^2} \approx -13$, can be written in analytical form:
\BEA
c^{\rm bos,2L}_L & = & 
\frac{1}{30}\left[107+23(1-4\sw^2)^2\right]
\approx 3.6,
\label{resSMbosonic2}
\EEA
in agreement with \citere{g-2SM2lA}. The coefficient 
$c^{\rm bos,2L}_0$ has a more involved analytical form. 
In \reffi{fig:SMbosonic} we show the results for $\amu^{\rm bos,2L}$
from \refeq{resSMbosonic} as a function of $\MHSM$. The variation
comes from the non-logarithmic piece~$c_0^{\rm bos,2L}$. The size of
the (constant) logarithmic contributions is also indicated. It can be
seen that the variation with $\MHSM$ is at the level of $0.3 \times 10^{-10}$
for $100 \gev \lsim \MHSM \lsim 500 \gev$.
These numerical values also confirm the results of
\citere{g-2SM2lA}.
The logarithmic piece in
(\ref{resSMbosonic}) alone is already a very good approximation
and leads to 
\BEA
\amu^{\rm bos,2L} & = & (-2.2\pm0.2) \times 10^{-10}~,
\EEA
corresponding to 
a reduction of the electroweak one-loop contribution by about $11\%$.


\section{Two-Higgs-doublet contributions}

\label{sec:2hdm}

In the MSSM, the bosonic electroweak two-loop contributions are
different compared to the SM because of the extended MSSM Higgs
sector. In this section we present the result for the
contributions of this class, defined by selecting all MSSM two-loop
diagrams without a closed loop of fermions or sfermions and without
pure QED-diagrams,
see \reffi{fig:logMHdiag}. 

\input{EWdiagrams.tex}

The result $\amuSUbos$ reads
\BEA
\amuSUbos & = & 
\frac{5}{3}\frac{\gf m_\mu^2}{8 \pi^2\sqrt{2}}
 \;\frac{\al}{\pi}
\KL c^\SUbos_L \log\frac{m_\mu^2}{\MW^2} 
+ c^\SUbos_0 \KR
\label{resTHDMbosonic}
\EEA
where the coefficient of the logarithm is given by
\BEA
\label{cLTHDM}
c^\SUbos_L & = & 
\frac{1}{30}\left[98+9c_L^{h}+23(1-4\sw^2)^2\right],\\
\label{cLh}
c_L^{h} & = & \frac{c_{2\be} \MZ^2}{c_\be} 
\KKL \frac{c_\al c_{\al+\be}}{\mH^2} + \frac{s_\al s_{\al+\be}}{\mh^2} \KKR~.
\EEA
Here $\be$ is defined by the ratio of the two Higgs-vacuum
expectation values, $\tb=v_2/v_1$; $m_{h,H}$ and $\al$ are
the masses and the mixing angle in the $\cp$-even Higgs sector, and
$c_\al\equiv \cos\al$, etc. 
As we will discuss below,  $c_L^h = 1$, and thus the logarithms in the
SM and the MSSM are identical.

\begin{figure}[htb!]
\begin{center}
\epsfig{figure=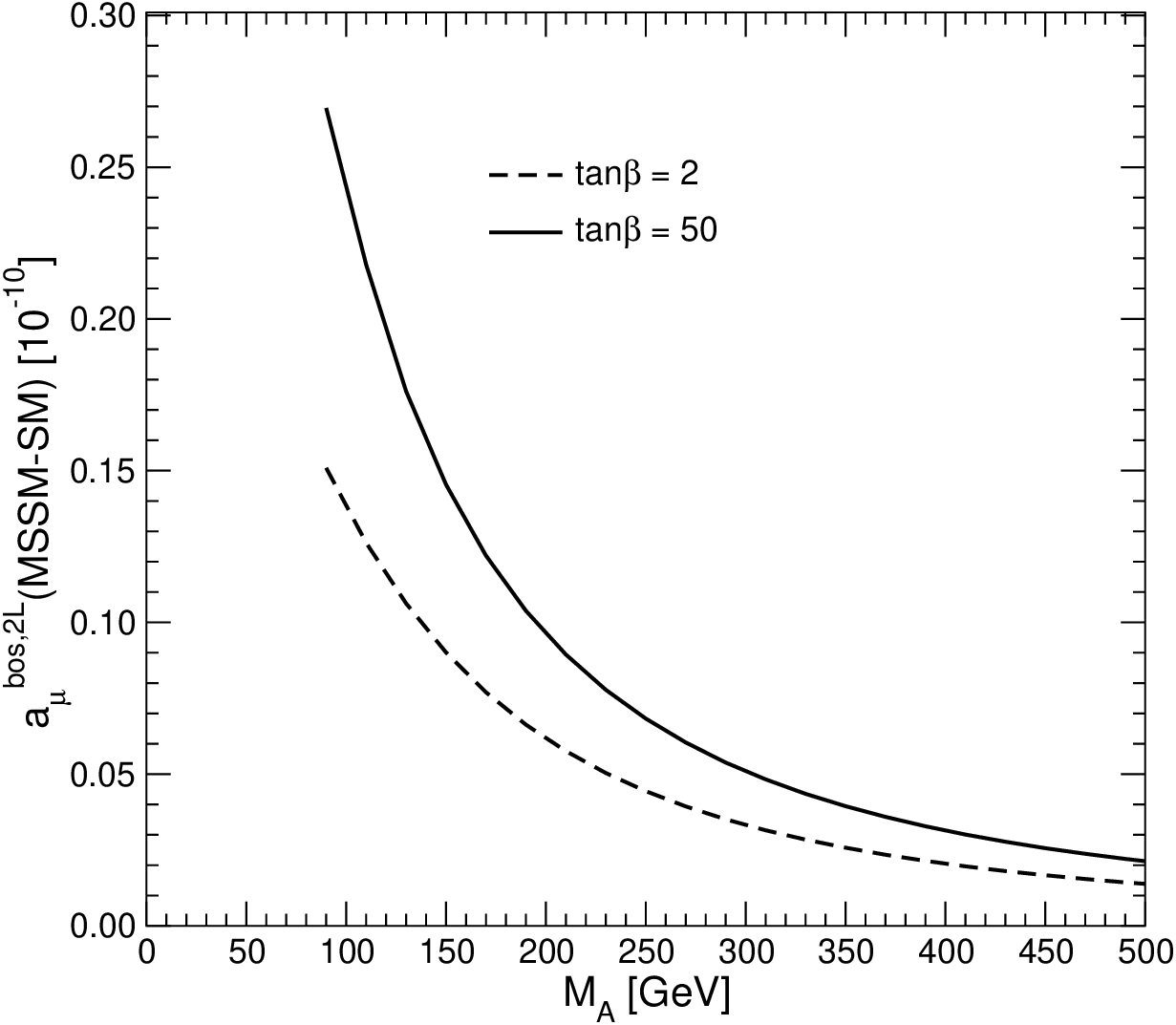,width=12cm,height=8cm}
\caption{%
The difference of $\amu^{\rm bos,2L}$ in the MSSM and the SM is shown
as a function of $\MA$. The SM Higgs-boson mass has been set equal to the
light $\cp$-even MSSM Higgs mass. $\tb$ has been set to $\tb = 2,50$.
The differences between the MSSM and the SM come solely from the
different coefficients $c_0^{\SUbos}$ and $c_0^{\rm bos,2L}$.
}
\label{fig:MAamuMSSM-SM}
\end{center}
\end{figure}

Let us first comment on the non-logarithmic contribution in
(\ref{resTHDMbosonic}). 
The non-logarithmic coefficient $c^\SUbos_0$ is a
complicated function of the full MSSM Higgs sector, which is
determined by two parameters, $\tb$ and the pseudo-scalar
Higgs-boson mass $\MA$. However, in \reffi{fig:MAamuMSSM-SM} it is
demonstrated that the full results for $\amuSUbos$ can be quite well
approximated by the corresponding SM 
quantity if $\MHSM$ is set equal to the light $\cp$-even Higgs-boson mass
of the MSSM, $\MHSM = \mh$, 
except in parameter scenarios with extremely light $\cp$-odd
Higgs-boson mass. For $\MA \gsim 100 \gev$ the difference stays below
$0.3 \times 10^{-10}$. 
The difference between the MSSM and the SM comes solely from the
different coefficients $c_0^{\SUbos}$ and $c_0^{\rm bos,2L}$.

As in the SM, the logarithmic piece in (\ref{resTHDMbosonic}) is an
excellent approximation of the full bosonic result. At first sight,
the coefficient (\ref{cLTHDM}) seems different from the corresponding
SM one, owing to the appearance of Higgs-mass dependent
terms. However, the MSSM Higgs-masses satisfy the relation
\BEA
\frac{c_{2\be} \MZ^2}{c_\be} 
\KKL \frac{c_\al c_{\al+\be}}{\mH^2} + \frac{s_\al s_{\al+\be}}{\mh^2} \KKR
& = & 1
\label{chrelation}
\EEA
at tree-level. Hence, the coefficient $c_L^h=1$ in (\ref{cLTHDM}), and
the logarithmic pieces of the MSSM and SM bosonic two-loop corrections
are identical as mentioned above.
This finding can be explained in several different ways. One
way is to apply the analysis of the
$\log\frac{m_\mu^2}{\MW^2}$-terms performed in \citere{g-2MSSMlog2l}
to the bosonic contributions. This shows 
that they must be identical in the SM and MSSM since the corresponding
bosonic one-loop diagrams are identical. The latter statement is true
since the additional MSSM Higgs bosons do not appear at the one-loop
level, in particular since there is no $\ga W^\pm
H^\mp$-coupling.%
\footnote{
More precisely, the additional MSSM Higgs
  bosons as well as the SM Higgs boson do appear in one-loop diagrams,
  but their contributions are suppressed by two additional muon Yukawa
  couplings and hence negligible.
}%

Alternatively, one can directly investigate the first Feynman diagram in
\reffi{fig:logMHdiag} (with $\phi = h,H$ and $\psi = G^\pm$) in order
to understand how the logarithms in the 
SM and MSSM emerge. 
The diagrams with photon and physical Higgs exchange and an inner
$G^\pm-W^\pm$ loop are responsible for
the Higgs-mass dependence in the 
coefficients of $\log\frac{m_\mu^2}{\MW^2}$. In the corresponding SM
diagram (with $\phi = H^{\rm SM}$) the $G^+ G^-H^{\rm SM}$-coupling 
is proportional to $\MHSM^2/\MW$, and the
dependence on $\MHSM$ exactly cancels with the suppression
coming from the Higgs propagator. In the MSSM the couplings of
$G^\pm$ to physical Higgs bosons are given by gauge couplings, hence
leading to the structure of $c_L^h$ shown above in \refeq{cLh} 
and to the seeming $m_{h,H}$-dependence of the logarithm. 

However, as the appearance of
Goldstone bosons signals, such Feynman diagrams are not gauge
independent by themselves. For example, in the 
unitary gauge there are no Goldstone
bosons and such diagrams do not exist. Similarly, in the background
field gauge \cite{BFM} or in the nonlinear gauge used in
\citere{g-2SM2lA} there is no $\ga W^\pm G^\mp$-vertex, and again the
diagrams in Fig.\ \ref{fig:logMHdiag} do not exist. In all these
gauges there are no diagrams with virtual photon and physical Higgs,
hence it is manifest that the logarithms in the SM and MSSM are
identical.

In the Feynman gauge, which we we have chosen,
this fact is not immediately
obvious but follows from relation (\ref{chrelation}), which in turn is
a consequence of global gauge invariance. In fact, one can 
easily derive that the Higgs mass matrix $\Ga_{\phi_i\phi_j}$
(corresponding to the denominators in (\ref{chrelation})), where
$\phi_{i,j}$ are the $\cp$-even interaction Higgs eigenstates, is
related to the couplings $\Ga_{\phi_i G^+G^-}$ (corresponding to the
numerators in~(\ref{chrelation})).%
\footnote{
In \citere{TB}, eqs.\ (27,28), a similar relation between
  the Higgs mass matrix $\Ga_{\phi_i\phi_j}$ and the gauge
  dependence of tadpole diagrams has been derived. Gauge dependent
  tadpole diagrams are in particular the ones with $G^\pm$~loops, thus
  involving the coupling $\Ga_{\phi_i G^+G^-}$, so that this
  constitutes again a gauge relation between $\Ga_{\phi_i\phi_j}$ and 
$\Ga_{\phi_i G^+G^-}$.
}%

This discussion has a practical implication concerning the treatment of
higher-order contributions to the Higgs-boson masses. In other
sectors like the sfermion-loop or chargino/neutralino-loop sector,
leading three-loop effects can be taken into account by using
the loop-corrected values for the Higgs-boson masses $M_{h,H}$ in the
propagators. In the present sector, however, using loop-corrected
Higgs-boson masses would spoil the validity of (\ref{chrelation}),
which is required by gauge invariance. Three-loop effects in the Higgs
propagators would be taken into account but three-loop effects in the
couplings $\Ga_{\phi_i G^+G^-}$ would be neglected, resulting in fake
large effects. Therefore, tree-level Higgs-boson masses have to be
used in the propagators for the bosonic two-loop contributions (as has
been done in \reffi{fig:MAamuMSSM-SM}).

The shift induced by replacing the tree-level Higgs-boson masses by 
loop corrected masses in the two-loop contributions gives an indication
of the possible size of bosonic three-loop corrections.
A shift of \order{50 \gev} in $\mh$ induces a
variation in $\amu$ of up to about $0.2 \times 10^{-10}$ as shown in
\reffi{fig:SMbosonic}, which can be regarded as a three-loop
uncertainty.


\section{Chargino/neutralino loop contributions}

The 2-loop contributions to $\amu$ containing a closed
chargino/neutralino loop constitute a separately UV-finite and
gauge-independent class. The corresponding diagrams are shown in 
\reffi{fig:CNdiagrams}. In this section we discuss the numerical
impact and the parameter dependence of this class.

\input ChaNeuDiagrams

The chargino/neutralino two-loop contributions, $\amuX$,
depend on the mass parameters for the charginos and neutralinos $\mu$,
$M_{1,2}$, the $\cp$-odd Higgs mass $\MA$, and $\tb$. Since we use
the loop-corrected values for the $\cp$-even Higgs masses
$M_{h,H}$~\cite{mhiggslong,mhiggsAEC,feynhiggs} 
in the propagators for this class of diagrams,
formally an effect of 3-loop order, $\amuX$ 
also has a slight dependence on other MSSM parameters through $M_{h,H}$
(we denote the loop-corrected masses as $M_{h,H}$ and the tree-level
masses as $m_{h,H}$; see also the discussion in \refse{sec:2hdm}).
It is interesting to
note that, contrary to \citere{delrho2lC},  no tree-level relations in
the Higgs sector were needed in order to find a UV-finite result.
This is due to the fact that each two-loop diagram contributing to
$(g-2)_\mu$ together with its corresponding subloop renormalization is
finite. 

\begin{figure}[htb!]
\begin{center}
\epsfig{figure=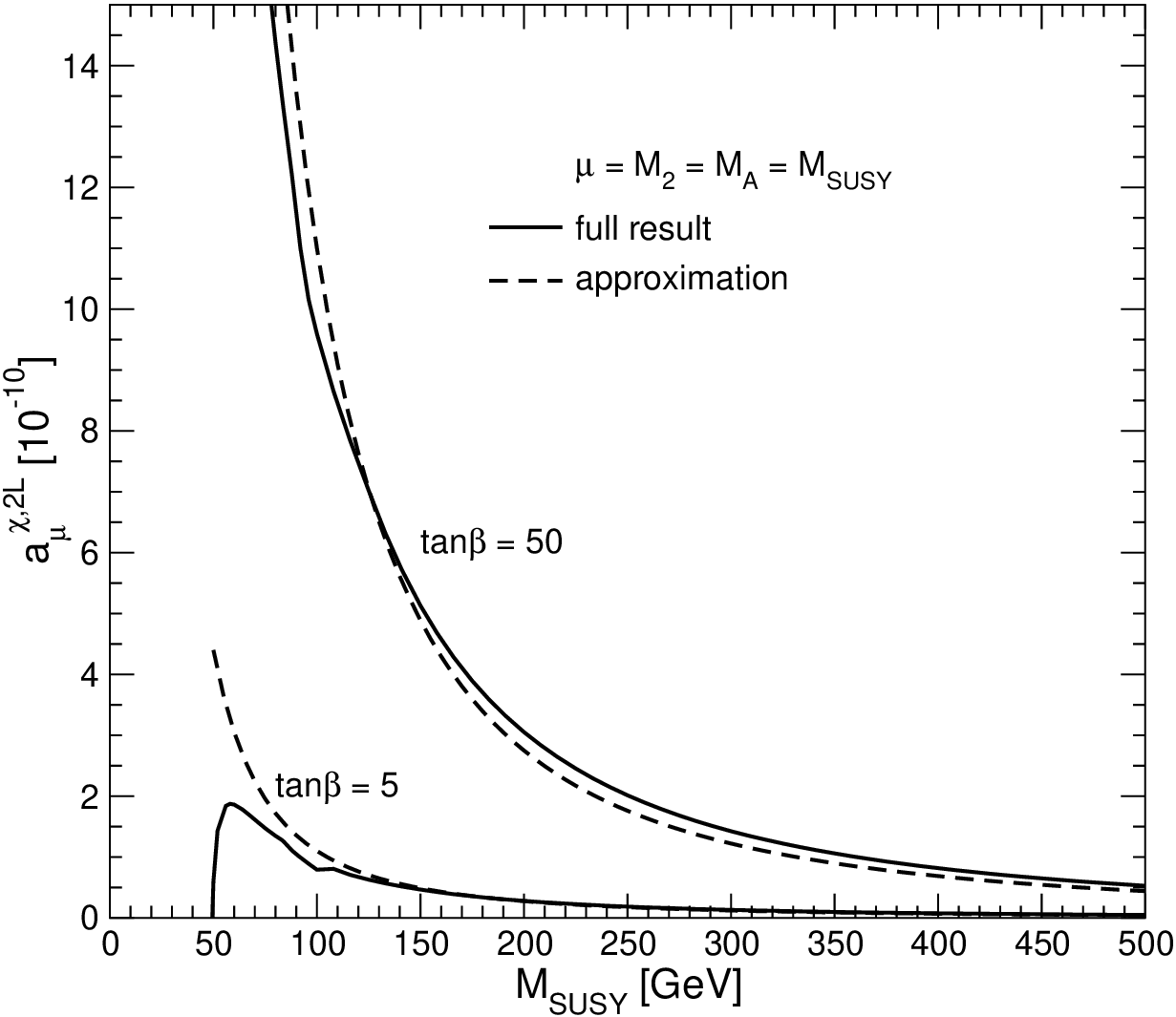,width=12cm,height=8cm}
\caption{%
Comparison of the full result for $\amuX$ with the approximation
(\ref{chaneuapprox}) for $\mu=M_2=\MA \equiv \msusy$. The full
lines show the complete result, while the dashed lines indicate the 
approximation. For
the upper curves $\tb=50$ has been used, for the lower ones
$\tb=5$.}
\label{fig:CNapprox}
\end{center}
\end{figure}

It turns out that the parameter dependence of $\amuX$ is quite
straightforward. If all supersymmetric mass scales are set equal,%
\footnote{
For simplicity we assume throughout this
paper that $M_{1,2}$ are related by the GUT relation
$M_1=5/3\;\sw^2/\cw^2 \; M_2$.
}
$\mu=M_2=\MA\equiv \msusy$, the approximate leading behaviour of
$\amuX$ is simply given by $\tb/\msusy^2$, and we 
find the approximate relation
\BE
\amuX \approx 11\times10^{-10}
\left(\frac{\tb}{50}\right)
\left(\frac{100 \gev}{M_{\rm SUSY}}\right)^2\;{\mbox{sign}}(\mu)~.
\label{chaneuapprox}
\end{equation}
As shown in \reffi{fig:CNapprox}, the approximation is very
good except for very small $\msusy$ and small $\tb$, where
the leading term is suppressed by the small $\mu$, 
and subleading terms begin to dominate.

%
\begin{figure}[hb!]
\vspace{-1.5cm}
\begin{center}
\unitlength=1.25bp
\epsfig{figure=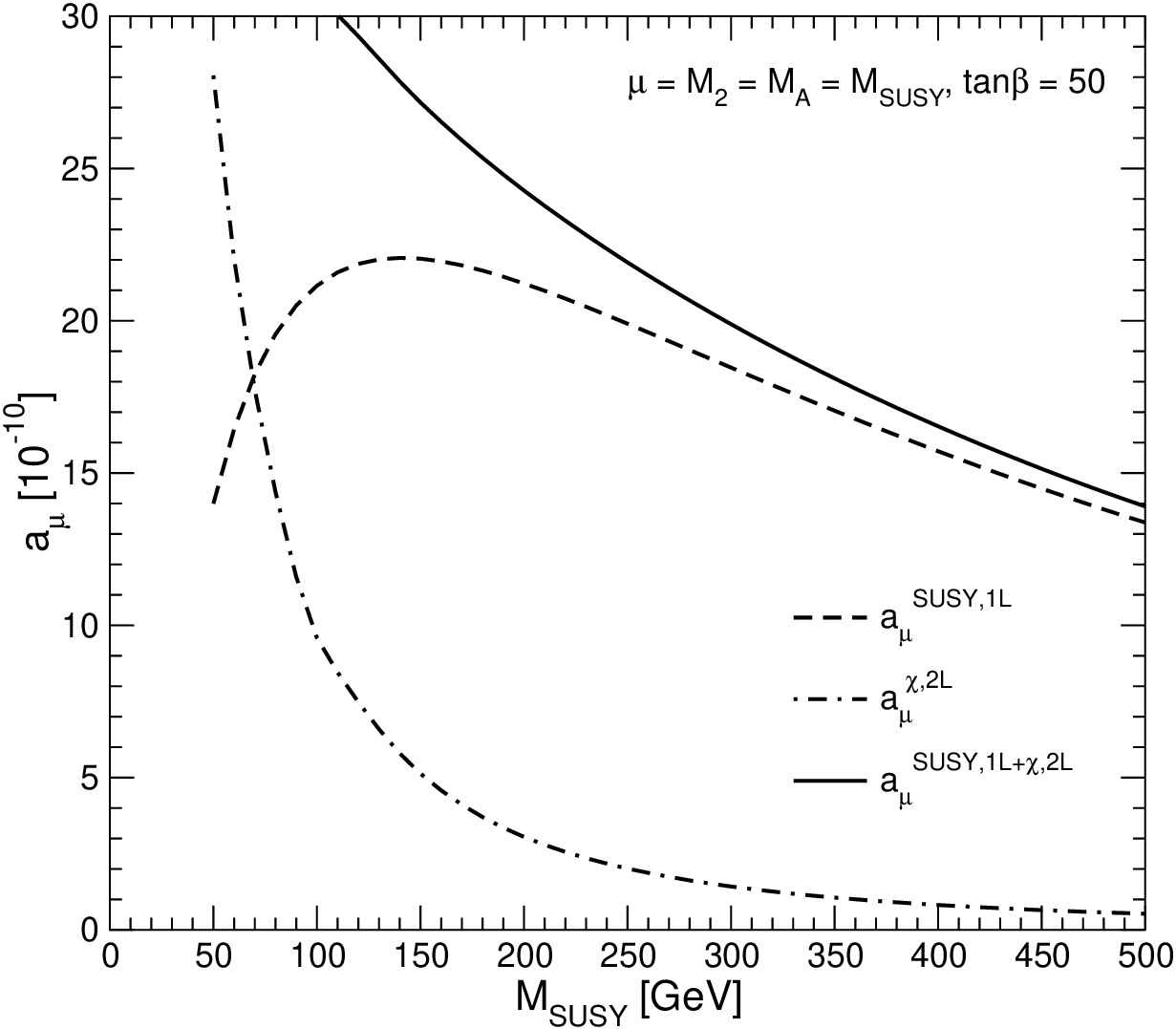,width=12cm,height=8cm}
\caption{Comparison of the supersymmetric one-loop result $\amu^{\rm
    SUSY,1L}$ (dashed) with the two-loop chargino/neutralino contributions
    $\amuX$ (dash-dotted) and the sum (full line). The parameters are
    $\mu=M_2=\MA\equiv \msusy$, 
    $\tb=50$, and the sfermion mass parameters are set to 1TeV.}
\label{fig:chaneueqmassplot}
\end{center}
\end{figure}

The chargino/neutralino sector does not only contribute to
$\amuX$ but already to $\amuSUoL$, so it is
interesting to compare the one- and two-loop contributions. 
For the case that all masses, including the smuon and
sneutrino masses, are set equal to $\msusy$, the one-loop and
two-loop contributions can be trivially compared using eqs.\
(\ref{susy1loop}), (\ref{chaneuapprox}), showing that the two-loop
contribution shifts the one-loop result by about~2\%. 

However, the chargino/neutralino sector might very well be significantly
lighter than the slepton sector of the second generation, in
particular in the light of FCNC and $\cp$-violating constraints, which
are more easily satisfied for heavy 1st and 2nd generation sfermions.
In \reffi{fig:chaneueqmassplot} the chargino/neutralino two-loop
contributions are therefore compared with the supersymmetric one-loop
contribution $\amuSUoL$ at fixed high smuon and sneutrino
masses $\Msl = 1 \tev$.
The other masses are again set equal, 
$\mu=M_2=\MA\equiv \msusy$. Furthermore, 
we use a large $\tb$ value, $\tb = 50$, which enhances the SUSY
contributions to $\amu$. 

We find that for $\msusy\lsim400 \gev$ the
two-loop contributions become more and more important. For $M_{\rm
  SUSY}\approx100 \gev$ they even amount to 50\% of the one-loop
contributions, which are suppressed by the large smuon and sneutrino
masses.

It is noteworthy that in the numerical analysis of the sfermion loop
contributions in \citere{g-2FSf} it was crucial to take into
account the experimental constraints on
the light Higgs-boson mass
$\Mh$~\cite{mhiggslong,mhiggsAEC,LEPHiggsSM,LEPHiggsMSSM},
electroweak precision
observables~\cite{delrhosusy1loop,delrhosusy2loop,delrho2lC} and the 
$b$-decays $B\to X_s\ga$ and $B_s\to\mu^+\mu^-$~\cite{bsmumu,bsg}.
In the case of chargino/neutralino loop contributions, however,
these experimental constraints have only little impact on the possible
choices of $\mu$, $M_2$ and $\MA$ since the
relevant observables depend strongly on the 3rd generation sfermion
mass parameters.

{}From \reffis{fig:CNapprox}, \ref{fig:chaneueqmassplot} one can read off
that the chargino/neutralino two-loop contributions can amount to about
$\pm 10 \times 10^{-10}$ if $\mu$, $M_2$, and $\MA$ are around 
$100 \gev$. 
This parameter region is rather constrained but not entirely excluded by
the experimental bounds from direct searches for charginos and
neutralinos, from Higgs searches and from $b$-physics.

\begin{figure}[htb!]
\vspace{2em}
\psfrag{MSl750TB25}{\footnotesize \hspace{-1em}$\Msl = 750 \gev, \tb = 25$}
\psfrag{MSl1000TB25}{\footnotesize \hspace{-1em}$\Msl = 1000 \gev, \tb = 25$}
\psfrag{MSl1000TB50}{\footnotesize \hspace{-1em}$\Msl = 1000 \gev, \tb = 50$}
\psfrag{MSl1500TB50}{\footnotesize \hspace{-1.7em}$\Msl = 1500 \gev, \tb = 50$}
\begin{center}
\begin{picture}(440,440)\epsfxsize=6.5cm
 \put(000,000){\epsfbox{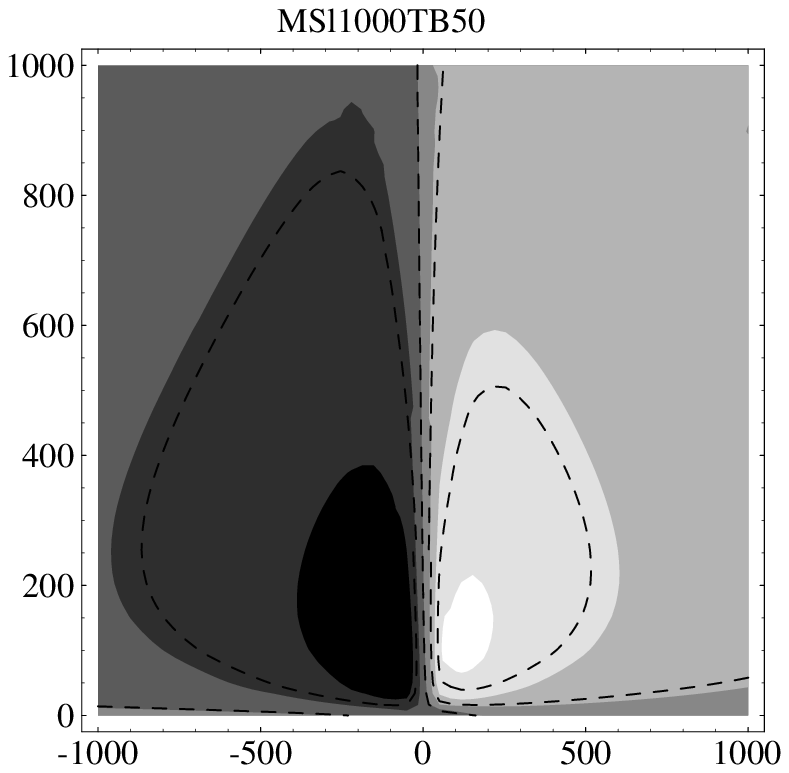}}
 \put(220,000){\epsfbox{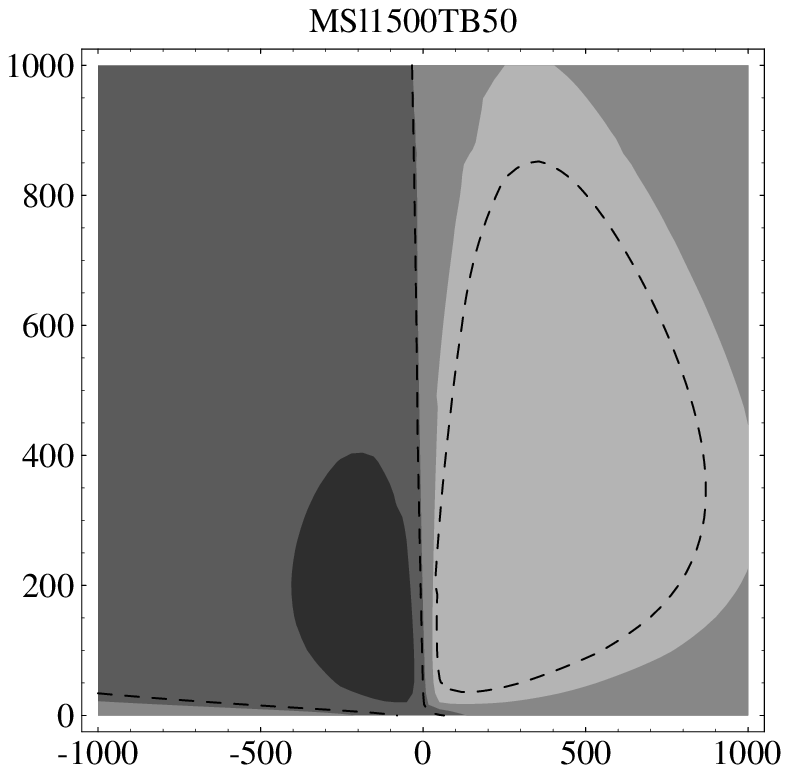}}
 \put(000,220){\epsfbox{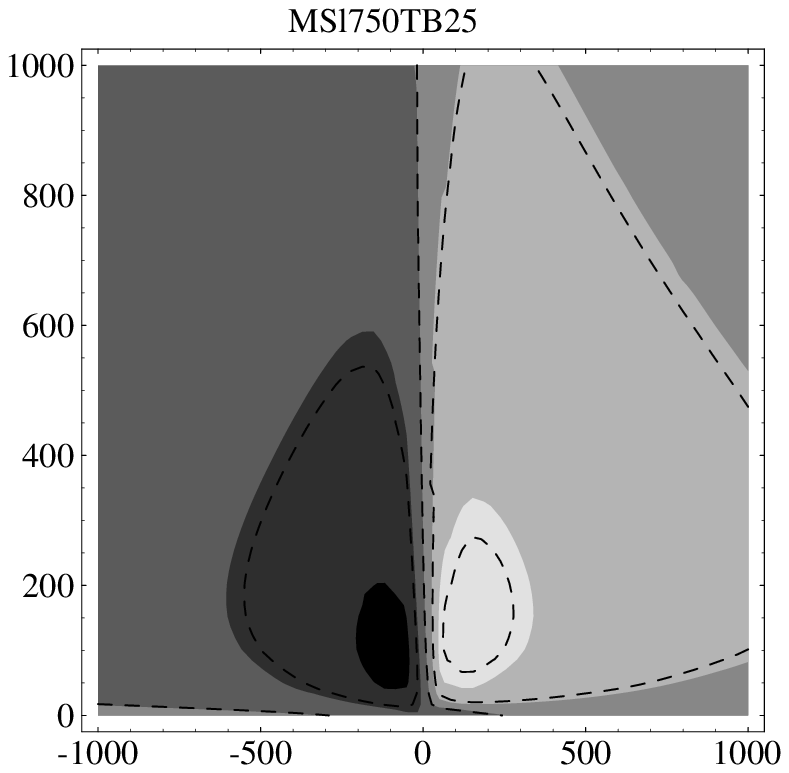}}
 \put(220,220){\epsfbox{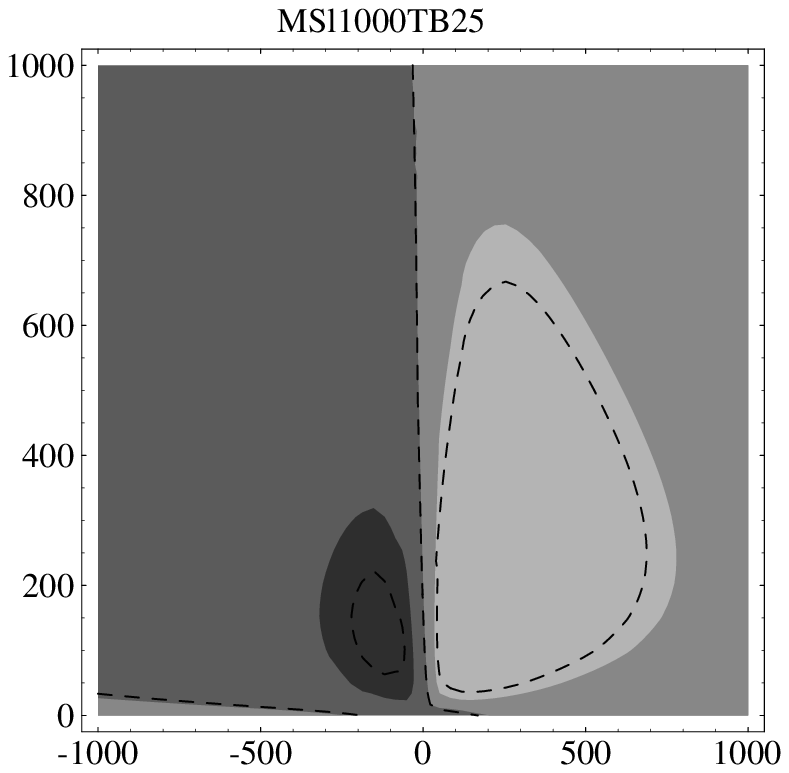}}
\put(005,180){\footnotesize $M_2$ [GeV]}
\put(225,180){\footnotesize $M_2$ [GeV]}
\put(005,400){\footnotesize $M_2$ [GeV]}
\put(225,400){\footnotesize $M_2$ [GeV]}
\put(150, -5){\footnotesize $\mu$ [GeV]}
\put(370, -5){\footnotesize $\mu$ [GeV]}
\put(150,215){\footnotesize $\mu$ [GeV]}
\put(370,215){\footnotesize $\mu$ [GeV]}
\put(110,060){\footnotesize $<1\si$}
\put(130,140){\footnotesize $1-2\si$}
\put(040,150){\footnotesize $3-4\si$}
%
\put(335,070){\footnotesize $1-2\si$}
\put(260,150){\footnotesize $3-4\si$}
\put(105,260){\scriptsize $<1\si$}
\put(120,300){\footnotesize $1-2\si$}
\put(040,370){\footnotesize $3-4\si$}
%
\put(330,300){\footnotesize $1-2\si$}
\put(350,370){\footnotesize $2-3\si$}
\put(260,370){\footnotesize $3-4\si$}
\end{picture}
\vspace{.2em}
\caption{Contours of $\amuSUoL+\amuX$ (solid border) and
  $\amuSUoL$ alone (dashed line) in the $\mu$--$M_2$-plane for 
$\MA = 200 \gev$. The slepton mass scale (which enters only the
  one-loop prediction) and $\tb$ are indicated for each plot.
  The contours are at $(24.5,15.5,6.5,-2.5,-11.5,-20.5)\times10^{-10}$ 
  corresponding to the central value of 
  $\amuexp-\amu^{\rm theo,SM} = (24.5 \pm 9.0)\times10^{-10}$ 
  and intervals of 1--5$\sigma$. 
}
\label{fig:mum2contour}
\end{center}
\vspace{2em}
\end{figure}

The dependence of $\amuSUoL+\amuX$ and $\amuSUoL$ on 
$\mu$ and $M_2$ is shown in the contour plots of
\reffi{fig:mum2contour}. We fix $\MA=200 \gev$ and vary $\tb$ and
the common smuon and sneutrino mass $\Msl$, which has an impact only
on the one-loop contribution. We have checked
  that these parameter choices are allowed essentially in the entire
  $\mu$--$M_2$-plane by the current experimental
  constraints mentioned above, provided the $\Stop$~and $\Sbot$~mass
  parameters are of \order{1 \tev}. 
The contours drawn in 
\reffi{fig:mum2contour} correspond to the $1\si$, $2\si$, \ldots regions
around the value 
$\amuexp - \amu^{\rm theo,SM}=(24.5\pm9.0)\times10^{-10}$, 
based on \citeres{g-2HMNT,LBLnew}. 
We find that for the investigated parameter space the SUSY prediction
for $\amu$ lies mostly in the $0-2\,\si$ region if $\mu$ is positive.
However, the new two-loop corrections shift the $1\,\si$ and $2\,\si$
contours considerably. This effect is more pronounced for smaller
$\tb$ and larger~$\Msl$.

\begin{figure}[htb]
  \begin{center}
\epsfig{figure=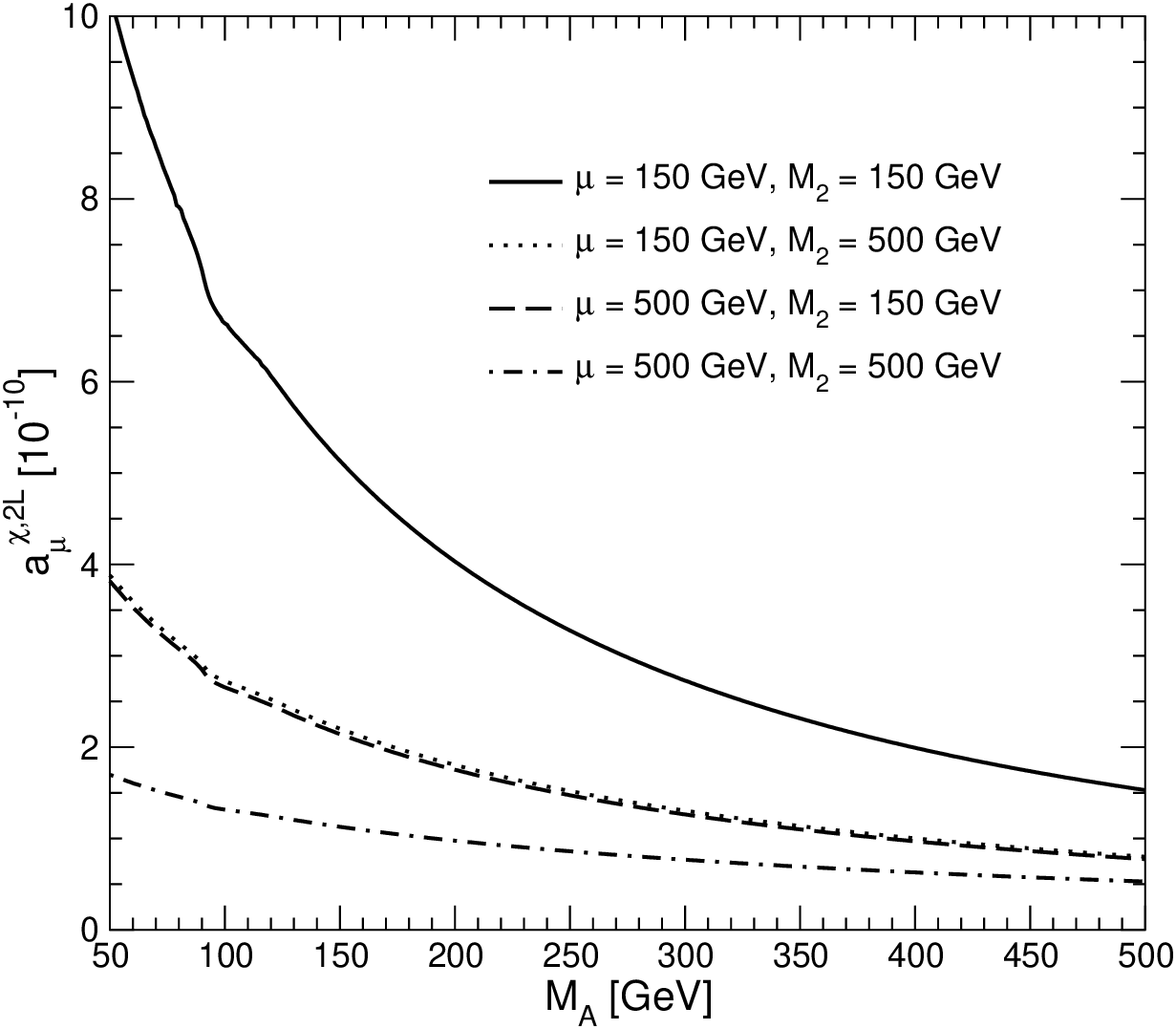,width=12cm,height=8cm}
\caption{Dependence of $\amuX$ on the $\cp$-odd Higgs mass $\MA$ for
  $\tb=50$ and  $\mu=M_2=150,500 \gev$.}
\label{fig:chaneuMAdep}
  \end{center}
\end{figure}

The $\MA$-dependence of the pure two-loop contributions $\amuX$
is shown in \reffi{fig:chaneuMAdep} for $\mu$, $M_2=150$,
$500 \gev$. For the smaller values of $\mu$ and $M_2$, $\amuX$
behaves approximately as $1/\MA$; for the larger values  of $\mu$ and
$M_2$ the $\MA$-dependence is less pronounced.

\section{Conclusions}

We have calculated two kinds of two-loop contributions to $(g-2)_\mu$:
the purely bosonic two-loop corrections (diagrams involving $\mu$,
$\nu_\mu$, gauge and Higgs bosons) in the SM and the MSSM, and the 
two-loop corrections involving a neutralino/chargino subloop in the
MSSM. 

In the SM our calculation provides an independent check and a slight
extension of \citeres{g-2SM2lA,g-2SM2lB}, which was up to now the only
calculation of the bosonic two-loop corrections, using the
approximation $\MHSM \gg \MW$. We find good agreement with
\citeres{g-2SM2lA,g-2SM2lB}. 
The variation with $\MHSM$ is at the level of $0.3 \times 10^{-10}$
for $100 \gev \lsim \MHSM$ $\lsim 500 \gev$.
The final result can be well approximated by the $\MHSM$-independent
logarithmic piece, \refeq{resSMbosonic2}, 
$\amu^{\rm bos,2L} = (-2.2 \pm 0.2) \times 10^{-10}$,
where the central value corresponds to $\MHSM \approx 200 \gev$. 

In the MSSM our calculation completes the program begun in
\citere{g-2FSf} to calculate all MSSM two-loop corrections to 
two-Higgs-doublet model
one-loop diagrams. This class of corrections has a potentially larger
impact on the MSSM prediction of $(g-2)_\mu$ than the higher-order
corrections to purely SUSY one-loop diagrams, since it is not
suppressed in case the smuon and sneutrino masses are large. In addition 
to the SM fermion and scalar fermion loop corrections 
derived in \citere{g-2FSf}, we have
calculated here the diagrams with a neutralino/chargino loop as well
as the purely bosonic two-loop diagrams in the MSSM (i.e.\ diagrams
involving no SUSY particles). 

For the purely bosonic two-loop corrections in the MSSM we find only a small
deviation from the SM result. The logarithmic piece is identical,
while the non-logarithmic piece causes only a difference in $\amu$ 
below the level of $0.3 \times 10^{-10}$ compared to the SM result 
with a light Higgs. The result for the purely bosonic two-loop
corrections in the MSSM lies in the range 
\mbox{
$-1.5 \times 10^{-10} \lsim \amuSUbos \lsim -2.0 \times 10^{-10}$},
depending on the values of $\MA$ and $\tb$. 

A sizable shift, on the other hand, arises from the two-loop contributions 
where a neutralino/chargino loop is inserted into a two-Higgs-doublet model 
one-loop diagram. Depending on the values of $\mu$,
$M_2$, $\MA$, and $\tb$, the result can amount up to $\pm 10 \times 10^{-10}$. 
For example, if all masses are set equal to a common scale $\msusy$ we
find the approximate relation 
\BE
\amuX \approx 11\times10^{-10}
\left(\frac{\tb}{50}\right)
\left(\frac{100 \gev}{M_{\rm SUSY}}\right)^2\;{\mbox{sign}}(\mu)~.
\end{equation}
Thus, the regions in the MSSM parameter space which satisfy the
requirement that $\amu$ agrees with the experimental value at the
$2\,\si$ level can be shifted significantly
as a consequence of the new two-loop corrections. 

The analytical results presented here will be implemented into the
Fortran code \fh\ (see: {\tt www.feynhiggs.de}).
Based on the new results presented in this paper, 
the last missing piece of the MSSM two-loop contributions to 
$(g-2)_\mu$ are the two-loop corrections to the SUSY one-loop diagrams.
Their inclusion, which will be desirable in order to further reduce the
theoretical uncertainty, will constitute the first complete two-loop
result within the MSSM.


\subsection*{Acknowledgements}

S.H.\ thanks the 
Institute for Particle Physics Phenomenology (IPPP), University of Durham,
for hospitality during the early stages of this work. 
D.S.\ thanks the CERN Theory division for hospitality when finalizing
this work.
G.W.\ thanks B.~Krause for interesting discussions in the early stages
of this work.




\end{document}

%% file: QEDdiagrams.tex
\begin{figure}[htb!]
\begin{center}
\unitlength=0.9bp%

\begin{feynartspicture}(290,160)(2,1)

\FADiagram{}
\FAProp(0.,15.)(5.,15.)(0.,){/Straight}{1}
\FALabel(2.5,13.93)[t]{$\mu$}
\FAProp(0.,5.)(5.,5.)(0.,){/Sine}{0}
\FALabel(2.5,3.93)[t]{$\gamma$}
\FAProp(20.,10.)(13.4,10.)(0.,){/Straight}{-1}
\FALabel(16.7,11.07)[b]{$\mu$}
\FAProp(5.,15.)(5.,5.)(0.,){/Straight}{1}
\FALabel(3.93,10.)[r]{$\mu$}
\FAProp(7.5,13.5)(5.,15.)(0.,){/Sine}{0}
\FALabel(6.56747,15.0991)[bl]{$\gamma$}
\FAProp(10.5,11.5)(7.5,13.5)(0.8,){/Straight}{-1}
\FALabel(10.1798,14.5097)[bl]{$f$}
\FAProp(10.5,11.5)(7.5,13.5)(-0.8,){/Straight}{1}
\FALabel(8.34132,10.572)[tr]{$\bar f$}
\FAProp(10.5,11.5)(13.4,10.)(0.,){/Sine}{0}
\FALabel(12.1821,11.6467)[bl]{$\gamma$}
\FAProp(5.,5.)(13.4,10.)(0.,){/Straight}{1}
\FALabel(9.5128,6.6481)[tl]{$\mu$}
\FAVert(10.5,11.5){0}
\FAVert(7.5,13.5){0}
\FAVert(5.,15.){0}
\FAVert(5.,5.){0}
\FAVert(13.4,10.){0}

\FADiagram{}
\FAProp(0.,15.)(5.,15.)(0.,){/Straight}{1}
\FALabel(2.5,16.07)[b]{$\mu$}
\FAProp(0.,5.)(5.,5.)(0.,){/Sine}{0}
\FALabel(2.5,3.93)[t]{$\gamma$}
\FAProp(20.,10.)(13.4,10.)(0.,){/Straight}{-1}
\FALabel(16.7,11.07)[b]{$\mu$}
\FAProp(9.,12.5)(5.,15.)(0.,){/Sine}{0}
\FALabel(7.3415,14.5844)[bl]{$\gamma$}
\FAProp(9.,12.5)(13.4,10.)(0.,){/Sine}{0}
\FALabel(11.4857,12.1177)[bl]{$\gamma$}
\FAProp(5.,15.)(5.,5.)(0.,){/Straight}{1}
\FALabel(3.93,10.)[r]{$\mu$}
\FAProp(5.,5.)(13.4,10.)(0.,){/Straight}{1}
\FALabel(9.5128,6.6481)[tl]{$\mu$}
\FAVert(5.,15.){0}
\FAVert(5.,5.){0}
\FAVert(13.4,10.){0}
\FAVert(9.,12.5){1}

\end{feynartspicture}

\vspace{-2.5em}
\caption{Generic QED diagrams with a fermion loop insertion (left) and
  the corresponding counterterm diagram (right).
The same kind of diagram exists with a scalar inserted instead of a
  fermion. 
}
\label{fig:QEDgeneral}
\end{center}
\end{figure}

%% file: EWdiagrams.tex
\begin{figure}[htb!]
\begin{center}
\unitlength=0.9bp%

\begin{feynartspicture}(432,140)(3,1)

\FADiagram{}
\FAProp(0.,15.)(5.,15.)(0.,){/Straight}{1}
\FALabel(2.5,16.07)[b]{$\mu$}
\FAProp(0.,5.)(5.,5.)(0.,){/Sine}{0}
\FALabel(2.5,3.93)[t]{$\gamma$}
\FAProp(20.,10.)(15.,10.)(0.,){/Straight}{-1}
\FALabel(17.5,11.07)[b]{$\mu$}
\FAProp(5.,15.)(15.,10.)(0.,){/Straight}{1}
\FALabel(10.2132,13.4064)[bl]{$\mu$}
\FAProp(5.,10.)(5.,15.)(0.,){/ScalarDash}{0}
\FALabel(4.18,12.5)[r]{$\phi$}
\FAProp(5.,10.)(5.,5.)(0.,){/ScalarDash}{-1}
\FALabel(3.93,7.5)[r]{$\psi$}
\FAProp(10.,7.5)(5.,10.)(0.,){/ScalarDash}{-1}
\FALabel(7.71318,9.65636)[bl]{$\psi$}
\FAProp(10.,7.5)(5.,5.)(0.,){/Sine}{1}
\FALabel(7.71318,5.34364)[tl]{$W$}
\FAProp(10.,7.5)(15.,10.)(0.,){/Sine}{0}
\FALabel(12.7132,7.84364)[tl]{$V$}
\FAVert(10.,7.5){0}
\FAVert(5.,10.){0}
\FAVert(5.,15.){0}
\FAVert(5.,5.){0}
\FAVert(15.,10.){0}

\FADiagram{}
\FAProp(0.,15.)(5.,15.)(0.,){/Straight}{1}
\FALabel(2.5,16.07)[b]{$\mu$}
\FAProp(0.,5.)(5.,5.)(0.,){/Sine}{0}
\FALabel(2.5,3.93)[t]{$\gamma$}
\FAProp(20.,10.)(15.,10.)(0.,){/Straight}{-1}
\FALabel(17.5,11.07)[b]{$\mu$}
\FAProp(5.,15.)(15.,10.)(0.,){/Straight}{1}
\FALabel(10.2132,13.4064)[bl]{$\mu$}
\FAProp(5.,10.)(5.,15.)(0.,){/ScalarDash}{0}
\FALabel(4.18,12.5)[r]{$\phi$}
\FAProp(5.,10.)(5.,5.)(0.,){/Sine}{-1}
\FALabel(3.93,7.5)[r]{$W$}
\FAProp(10.,7.5)(5.,10.)(0.,){/Sine}{-1}
\FALabel(7.71318,9.65636)[bl]{$W$}
\FAProp(10.,7.5)(5.,5.)(0.,){/Sine}{1}
\FALabel(7.71318,5.34364)[tl]{$W$}
\FAProp(10.,7.5)(15.,10.)(0.,){/Sine}{0}
\FALabel(12.7132,7.84364)[tl]{$V$}
\FAVert(10.,7.5){0}
\FAVert(5.,10.){0}
\FAVert(5.,15.){0}
\FAVert(5.,5.){0}
\FAVert(15.,10.){0}

\FADiagram{}
\FAProp(0.,15.)(5.,15.)(0.,){/Straight}{1}
\FALabel(2.5,16.07)[b]{$\mu$}
\FAProp(0.,5.)(5.,5.)(0.,){/Sine}{0}
\FALabel(2.5,3.93)[t]{$\gamma$}
\FAProp(20.,10.)(15.,10.)(0.,){/Straight}{-1}
\FALabel(17.5,11.07)[b]{$\mu$}
\FAProp(5.,15.)(15.,10.)(0.,){/Straight}{1}
\FALabel(10.2132,13.4064)[bl]{$\mu$}
\FAProp(5.,10.)(5.,15.)(0.,){/ScalarDash}{0}
\FALabel(4.18,12.5)[r]{$\phi$}
\FAProp(5.,10.)(5.,5.)(0.,){/GhostDash}{1}
\FALabel(3.93,7.5)[r]{$u_-$}
\FAProp(10.,7.5)(5.,10.)(0.,){/GhostDash}{1}
\FALabel(7.71318,9.65636)[bl]{$u_-$}
\FAProp(10.,7.5)(5.,5.)(0.,){/GhostDash}{-1}
\FALabel(7.71318,5.34364)[tl]{$u_-$}
\FAProp(10.,7.5)(15.,10.)(0.,){/Sine}{0}
\FALabel(12.7132,7.84364)[tl]{$V$}
\FAVert(10.,7.5){0}
\FAVert(5.,10.){0}
\FAVert(5.,15.){0}
\FAVert(5.,5.){0}
\FAVert(15.,10.){0}

\end{feynartspicture}
\vspace{-1.5em}


\begin{feynartspicture}(432,140)(3,1)

\FADiagram{}
\FAProp(0.,15.)(5.,15.)(0.,){/Straight}{1}
\FALabel(2.5,16.07)[b]{$\mu$}
\FAProp(0.,5.)(5.,5.)(0.,){/Sine}{0}
\FALabel(2.5,3.93)[t]{$\gamma$}
\FAProp(20.,10.)(15.,10.)(0.,){/Straight}{-1}
\FALabel(17.5,11.07)[b]{$\mu$}
\FAProp(5.,15.)(15.,10.)(0.,){/Straight}{1}
\FALabel(10.2132,13.4064)[bl]{$\mu$}
\FAProp(5.,10.5)(5.,15.)(0.,){/Sine}{0}
\FALabel(6.07,12.75)[l]{$V$}
\FAProp(5.,10.5)(5.,5.)(0.545455,){/ScalarDash}{-1}
\FALabel(2.43,7.75)[r]{$\psi$}
\FAProp(5.,10.5)(5.,5.)(-0.545455,){/ScalarDash}{1}
\FALabel(7.27,9.)[l]{$\psi$}
\FAProp(15.,10.)(5.,5.)(0.,){/Sine}{0}
\FALabel(10.2132,6.59364)[tl]{$V$}
\FAVert(5.,10.5){0}
\FAVert(5.,15.){0}
\FAVert(15.,10.){0}
\FAVert(5.,5.){0}

\FADiagram{}
\FAProp(0.,15.)(5.,15.)(0.,){/Straight}{1}
\FALabel(2.5,16.07)[b]{$\mu$}
\FAProp(0.,5.)(5.,5.)(0.,){/Sine}{0}
\FALabel(2.5,3.93)[t]{$\gamma$}
\FAProp(20.,10.)(15.,10.)(0.,){/Straight}{-1}
\FALabel(17.5,11.07)[b]{$\mu$}
\FAProp(5.,15.)(15.,10.)(0.,){/Straight}{1}
\FALabel(10.2132,13.4064)[bl]{$\mu$}
\FAProp(5.,15.)(7.5,9.)(0.,){/ScalarDash}{0}
\FALabel(5.48,12.98)[tr]{$\phi$}
\FAProp(5.,5.)(7.5,9.)(0.8,){/ScalarDash}{-1}
\FALabel(8.6844,5.6585)[tl]{$\psi$}
\FAProp(5.,5.)(7.5,9.)(-0.8,){/Sine}{1}
\FALabel(3.8156,8.3415)[br]{$W$}
\FAProp(15.,10.)(7.5,9.)(0.,){/Sine}{0}
\FALabel(11.4549,8.4436)[t]{$V$}
\FAVert(5.,15.){0}
\FAVert(5.,5.){0}
\FAVert(15.,10.){0}
\FAVert(7.5,9.){0}

\FADiagram{}
\FAProp(0.,15.)(5.,15.)(0.,){/Straight}{1}
\FALabel(2.5,13.93)[t]{$\mu$}
\FAProp(0.,5.)(5.,5.)(0.,){/Sine}{0}
\FALabel(2.5,3.93)[t]{$\gamma$}
\FAProp(20.,10.)(15.,10.)(0.,){/Straight}{-1}
\FALabel(17.5,11.07)[b]{$\mu$}
\FAProp(5.,15.)(5.,5.)(0.,){/Straight}{1}
\FALabel(3.93,10.)[r]{$\mu$}
\FAProp(5.,15.)(10.,12.5)(0.,){/Sine}{0}
\FALabel(6.6318,14.7436)[bl]{$\gamma$}
\FAProp(5.,5.)(15.,10.)(0.,){/Straight}{1}
\FALabel(10.2132,6.59364)[tl]{$\mu$}
\FAProp(15.,10.)(10.,12.5)(0.,){/Sine}{0}
\FALabel(13.1318,11.7436)[bl]{$Z$}
\FAProp(10.,12.5)(10.,12.5)(12.,16.){/ScalarDash}{-1}
\FALabel(12.289,16.8658)[bl]{$\psi$}
\FAVert(5.,15.){0}
\FAVert(5.,5.){0}
\FAVert(15.,10.){0}
\FAVert(10.,12.5){0}

\end{feynartspicture}
\vspace{-1.5em}


\begin{feynartspicture}(432,140)(3,1)

\FADiagram{}
\FAProp(0.,15.)(5.,15.)(0.,){/Straight}{1}
\FALabel(2.5,16.07)[b]{$\mu$}
\FAProp(0.,5.)(5.,5.)(0.,){/Sine}{0}
\FALabel(2.5,3.93)[t]{$\gamma$}
\FAProp(20.,10.)(15.,10.)(0.,){/Straight}{-1}
\FALabel(17.5,11.07)[b]{$\mu$}
\FAProp(5.,15.)(5.,5.)(0.,){/Straight}{1}
\FALabel(3.93,10.)[r]{$\mu$}
\FAProp(10.,12.5)(5.,15.)(0.,){/Sine}{0}
\FALabel(7.71318,14.6564)[bl]{$V$}
\FAProp(10.,12.5)(15.,10.)(0.,){/Straight}{1}
\FALabel(12.7132,12.1564)[bl]{$\mu$}
\FAProp(10.,7.5)(10.,12.5)(0.,){/Straight}{1}
\FALabel(8.93,10.)[r]{$\mu$}
\FAProp(10.,7.5)(5.,5.)(0.,){/Straight}{-1}
\FALabel(7.71318,5.34364)[tl]{$\mu$}
\FAProp(10.,7.5)(15.,10.)(0.,){/Sine}{0}
\FALabel(12.7132,7.84364)[tl]{$V$}
\FAVert(10.,7.5){0}
\FAVert(10.,12.5){0}
\FAVert(5.,15.){0}
\FAVert(5.,5.){0}
\FAVert(15.,10.){0}

\FADiagram{}
\FAProp(0.,15.)(5.,15.)(0.,){/Straight}{1}
\FALabel(2.5,16.07)[b]{$\mu$}
\FAProp(0.,5.)(5.,5.)(0.,){/Sine}{0}
\FALabel(2.5,3.93)[t]{$\gamma$}
\FAProp(20.,10.)(15.,10.)(0.,){/Straight}{-1}
\FALabel(17.5,11.07)[b]{$\mu$}
\FAProp(5.,15.)(15.,10.)(0.,){/Sine}{0}
\FALabel(10.2132,13.4064)[bl]{$V$}
\FAProp(5.,10.)(5.,15.)(0.,){/Straight}{-1}
\FALabel(3.93,12.5)[r]{$\mu$}
\FAProp(5.,10.)(5.,5.)(0.,){/Straight}{1}
\FALabel(3.93,7.5)[r]{$\mu$}
\FAProp(10.,7.5)(5.,10.)(0.,){/Sine}{0}
\FALabel(7.71318,9.65636)[bl]{$V$}
\FAProp(10.,7.5)(5.,5.)(0.,){/Straight}{-1}
\FALabel(7.71318,5.34364)[tl]{$\mu$}
\FAProp(10.,7.5)(15.,10.)(0.,){/Straight}{1}
\FALabel(12.7132,7.84364)[tl]{$\mu$}
\FAVert(10.,7.5){0}
\FAVert(5.,10.){0}
\FAVert(5.,15.){0}
\FAVert(5.,5.){0}
\FAVert(15.,10.){0}

\FADiagram{}
\FAProp(0.,15.)(5.,15.)(0.,){/Straight}{1}
\FALabel(2.5,16.07)[b]{$\mu$}
\FAProp(0.,5.)(5.,5.)(0.,){/Sine}{0}
\FALabel(2.5,3.93)[t]{$\gamma$}
\FAProp(20.,10.)(15.,10.)(0.,){/Straight}{-1}
\FALabel(17.5,11.07)[b]{$\mu$}
\FAProp(5.,15.)(15.,10.)(0.,){/Straight}{1}
\FALabel(10.2132,13.4064)[bl]{$\mu$}
\FAProp(5.,10.)(5.,15.)(0.,){/Sine}{0}
\FALabel(3.93,12.5)[r]{$Z$}
\FAProp(5.,10.)(5.,5.)(0.,){/ScalarDash}{1}
\FALabel(3.93,7.5)[r]{$\psi$}
\FAProp(10.,7.5)(5.,10.)(0.,){/ScalarDash}{1}
\FALabel(7.71318,9.65636)[bl]{$\psi$}
\FAProp(10.,7.5)(5.,5.)(0.,){/ScalarDash}{-1}
\FALabel(7.71318,5.34364)[tl]{$\psi$}
\FAProp(10.,7.5)(15.,10.)(0.,){/Sine}{0}
\FALabel(12.7132,7.84364)[tl]{$\gamma$}
\FAVert(10.,7.5){0}
\FAVert(5.,10.){0}
\FAVert(5.,15.){0}
\FAVert(5.,5.){0}
\FAVert(15.,10.){0}

\end{feynartspicture}


\vspace{-2.5em}
\caption{
Some two-Higgs-doublet model diagrams for $\amu$ with (depending on
the diagram) $\phi = h,H,A,G$; $\psi = G^\pm, H^\pm$; $V = \ga, Z$.
}
\label{fig:logMHdiag}
\end{center}
\end{figure}

%% file: ChaNeuDiagrams.tex
\begin{figure}[htb!]
\begin{center}
\unitlength=1.2bp%

\unitlength=0.9bp%
\begin{feynartspicture}(432,140)(3,1)

\FADiagram{}
\FAProp(0.,15.)(5.,15.)(0.,){/Straight}{1}
\FALabel(2.5,16.07)[b]{$\mu$}
\FAProp(0.,5.)(5.,5.)(0.,){/Sine}{0}
\FALabel(2.5,3.93)[t]{$\gamma$}
\FAProp(20.,10.)(13.4,10.)(0.,){/Straight}{-1}
\FALabel(16.7,11.07)[b]{$\mu$}
\FAProp(5.,15.)(13.4,10.)(0.,){/Straight}{0}
\FAProp(5.,11.5)(5.,15.)(0.,){/ScalarDash}{0}
\FALabel(4.18,13.25)[r]{$\phi$}
\FAProp(5.,8.5)(5.,11.5)(0.8,){/Straight}{0}
\FALabel(7.02,10.)[l]{$\tilde{\chi}$}
\FAProp(5.,8.5)(5.,11.5)(-0.8,){/Straight}{0}
\FAProp(5.,8.5)(5.,5.)(0.,){/ScalarDash}{0}
\FALabel(4.18,6.75)[r]{$\psi$}
\FAProp(5.,5.)(13.4,10.)(0.,){/Sine}{0}
\FALabel(9.5128,6.6481)[tl]{$V$}
\FAVert(5.,8.5){0}
\FAVert(5.,11.5){0}
\FAVert(5.,15.){0}
\FAVert(5.,5.){0}
\FAVert(13.4,10.){0}

\FADiagram{}
\FAProp(0.,15.)(5.,15.)(0.,){/Straight}{1}
\FALabel(2.5,16.07)[b]{$\mu$}
\FAProp(0.,5.)(5.,5.)(0.,){/Sine}{0}
\FALabel(2.5,3.93)[t]{$\gamma$}
\FAProp(20.,10.)(15.,10.)(0.,){/Straight}{-1}
\FALabel(17.5,11.07)[b]{$\mu$}
\FAProp(5.,15.)(5.,5.)(0.,){/ScalarDash}{0}
\FALabel(4.18,10.)[r]{$\psi$}
\FAProp(5.,15.)(15.,10.)(0.,){/Straight}{0}
\FAProp(8.,6.5)(5.,5.)(0.,){/Sine}{0}
\FALabel(6.71318,4.84364)[tl]{$V$}
\FAProp(12.,8.5)(8.,6.5)(0.8,){/Straight}{0}
\FALabel(9.09862,9.78276)[br]{$\tilde{\chi}$}
\FAProp(12.,8.5)(8.,6.5)(-0.8,){/Straight}{0}
\FAProp(12.,8.5)(15.,10.)(0.,){/Sine}{0}
\FALabel(13.7132,8.34364)[tl]{$V$}
\FAVert(12.,8.5){0}
\FAVert(8.,6.5){0}
\FAVert(5.,15.){0}
\FAVert(5.,5.){0}
\FAVert(15.,10.){0}

\FADiagram{}
\FAProp(0.,15.)(5.,15.)(0.,){/Straight}{1}
\FALabel(2.5,16.07)[b]{$\mu$}
\FAProp(0.,5.)(5.,5.)(0.,){/Sine}{0}
\FALabel(2.5,3.93)[t]{$\gamma$}
\FAProp(20.,10.)(15.,10.)(0.,){/Straight}{-1}
\FALabel(17.5,11.07)[b]{$\mu$}
\FAProp(5.,15.)(15.,10.)(0.,){/Straight}{0}
\FAProp(5.,10.)(5.,15.)(0.,){/ScalarDash}{0}
\FALabel(4.18,12.5)[r]{$\phi$}
\FAProp(5.,10.)(5.,5.)(0.,){/Straight}{0}
\FAProp(10.,7.5)(5.,10.)(0.,){/Straight}{0}
\FALabel(7.60138,9.43276)[bl]{$\tilde{\chi}$}
\FAProp(10.,7.5)(5.,5.)(0.,){/Straight}{0}
\FAProp(10.,7.5)(15.,10.)(0.,){/Sine}{0}
\FALabel(12.7132,7.84364)[tl]{$V$}
\FAVert(10.,7.5){0}
\FAVert(5.,10.){0}
\FAVert(5.,15.){0}
\FAVert(5.,5.){0}
\FAVert(15.,10.){0}

\end{feynartspicture}

\vspace{-1.5em}

\unitlength=0.9bp%
\begin{feynartspicture}(432,140)(3,1)

\FADiagram{}
\FAProp(0.,15.)(5.,15.)(0.,){/Straight}{1}
\FALabel(2.5,16.07)[b]{$\mu$}
\FAProp(0.,5.)(5.,5.)(0.,){/Sine}{0}
\FALabel(2.5,3.93)[t]{$\gamma$}
\FAProp(20.,10.)(13.4,10.)(0.,){/Straight}{-1}
\FALabel(16.7,11.07)[b]{$\mu$}
\FAProp(5.,15.)(13.4,10.)(0.,){/Straight}{0}
\FAProp(5.,11.5)(5.,15.)(0.,){/Sine}{0}
\FALabel(3.93,13.25)[r]{$V$}
\FAProp(5.,8.5)(5.,11.5)(0.8,){/Straight}{0}
\FALabel(7.02,10.)[l]{$\tilde{\chi}$}
\FAProp(5.,8.5)(5.,11.5)(-0.8,){/Straight}{0}
\FAProp(5.,8.5)(5.,5.)(0.,){/Sine}{0}
\FALabel(3.93,6.75)[r]{$V$}
\FAProp(5.,5.)(13.4,10.)(0.,){/Sine}{0}
\FALabel(9.5128,6.6481)[tl]{$V$}
\FAVert(5.,8.5){0}
\FAVert(5.,11.5){0}
\FAVert(5.,15.){0}
\FAVert(5.,5.){0}
\FAVert(13.4,10.){0}

\FADiagram{}
\FAProp(0.,15.)(5.,15.)(0.,){/Straight}{1}
\FALabel(2.5,16.07)[b]{$\mu$}
\FAProp(0.,5.)(5.,5.)(0.,){/Sine}{0}
\FALabel(2.5,3.93)[t]{$\gamma$}
\FAProp(20.,10.)(15.,10.)(0.,){/Straight}{-1}
\FALabel(17.5,11.07)[b]{$\mu$}
\FAProp(5.,15.)(5.,5.)(0.,){/Sine}{0}
\FALabel(3.93,10.)[r]{$V$}
\FAProp(5.,15.)(15.,10.)(0.,){/Straight}{0}
\FAProp(8.,6.5)(5.,5.)(0.,){/Sine}{0}
\FALabel(6.71318,4.84364)[tl]{$V$}
\FAProp(12.,8.5)(8.,6.5)(0.8,){/Straight}{0}
\FALabel(9.09862,9.78276)[br]{$\tilde{\chi}$}
\FAProp(12.,8.5)(8.,6.5)(-0.8,){/Straight}{0}
\FAProp(12.,8.5)(15.,10.)(0.,){/Sine}{0}
\FALabel(13.7132,8.34364)[tl]{$V$}
\FAVert(12.,8.5){0}
\FAVert(8.,6.5){0}
\FAVert(5.,15.){0}
\FAVert(5.,5.){0}
\FAVert(15.,10.){0}

\FADiagram{}
\FAProp(0.,15.)(5.,15.)(0.,){/Straight}{1}
\FALabel(2.5,16.07)[b]{$\mu$}
\FAProp(0.,5.)(5.,5.)(0.,){/Sine}{0}
\FALabel(2.5,3.93)[t]{$\gamma$}
\FAProp(20.,10.)(15.,10.)(0.,){/Straight}{-1}
\FALabel(17.5,11.07)[b]{$\mu$}
\FAProp(5.,15.)(15.,10.)(0.,){/Straight}{0}
\FAProp(5.,10.)(5.,15.)(0.,){/Sine}{0}
\FALabel(3.93,12.5)[r]{$V$}
\FAProp(5.,10.)(5.,5.)(0.,){/Straight}{0}
\FAProp(10.,7.5)(5.,10.)(0.,){/Straight}{0}
\FALabel(7.60138,9.43276)[bl]{$\tilde{\chi}$}
\FAProp(10.,7.5)(5.,5.)(0.,){/Straight}{0}
\FAProp(10.,7.5)(15.,10.)(0.,){/Sine}{0}
\FALabel(12.7132,7.84364)[tl]{$V$}
\FAVert(10.,7.5){0}
\FAVert(5.,10.){0}
\FAVert(5.,15.){0}
\FAVert(5.,5.){0}
\FAVert(15.,10.){0}

\end{feynartspicture}

\end{center}
\vspace{-2.5em}
\caption{Generic two-loop diagrams to $a_\mu$ with a closed
chargino/neutralino loop. $\phi$, $\psi$ denote the scalar particles
$h$, $H$, $A^0$, $H^\pm$ and $G^{0,\pm}$; $V$ denotes the vector
bosons $\gamma$, $Z$, $W^\pm$; $\tilde\chi$ stands for the
chargino/neutralino mass eigenstates $\tilde\chi^\pm_{1,2}$, 
$\tilde\chi^0_{1,2,3,4}$.}
\label{fig:CNdiagrams}
\end{figure}